\documentclass[letterpaper]{article} 
\usepackage{aaai2027}  
\usepackage[hyphens]{url}  
\usepackage{graphicx} 
\usepackage{natbib}  
\usepackage{caption} 
\usepackage{algorithm}
\usepackage{algorithmic}
\usepackage{amssymb}
\usepackage{amsmath}
\usepackage{dsfont}
\usepackage{subcaption}
\usepackage{appendix}

\usepackage{newfloat}
\usepackage{listings}
\DeclareCaptionStyle{ruled}{labelfont=normalfont,labelsep=colon,strut=off} 
\floatstyle{ruled}
\newfloat{listing}{tb}{lst}{}
\floatname{listing}{Listing}

\usepackage{booktabs}

\usepackage{multirow} 
\usepackage{subcaption}

\nocopyright

\title{AnyBand: Unified Multi-Bandwidth Speech Extension via Frequency-Aware In-Context Spectral Infilling}

\author{
    Junchuan Zhao\textsuperscript{\rm 1},
    Minh Duc Vu\textsuperscript{\rm 2},
    Bowen Zhang\textsuperscript{\rm 3},
    Ye Wang\textsuperscript{\rm 1}\corresponding
}

\affiliations{
    \textsuperscript{\rm 1}School of Computing, National University of Singapore\\
    \textsuperscript{\rm 2}Department of Statistics \& Data Science, National University of Singapore\\
    \textsuperscript{\rm 3}College of Computing and Data Science, Nanyang Technological University\\
    junchuan@u.nus.edu,
    minhduc.vu@u.nus.edu,
    bowen009@e.ntu.edu.sg,
    dcswangy@nus.edu.sg
}

\begin{document}

\maketitle

\begin{abstract}
Bandwidth extension (BWE) aims to recover missing high-frequency content from band-limited speech. Existing methods often formulate BWE as a fixed or predefined bandwidth conversion problem, potentially requiring cutoff-specific models or retraining when the input bandwidth changes. This assumption limits their applicability to practical scenarios where speech may arrive with diverse cutoff frequencies. We propose AnyBand, a unified BWE framework that recasts bandwidth extension as in-context spectral infilling. Motivated by prompt-based zero-shot speech generation, AnyBand conditions high-frequency generation on the observed low-frequency spectrum, using the available band as a frequency-domain prompt that conveys content, speaker, prosodic, and spectral-envelope cues. This formulation enables a single model to perform cutoff-conditioned generation over a continuous range of input bandwidths. AnyBand is trained with missing-band conditional flow matching and an Easy-to-Balanced cutoff curriculum over continuously sampled cutoff frequencies. To better exploit the spectral prompt, we introduce a frequency-aware Diffusion Transformer that models cross-frequency interactions and long-range temporal dependencies, followed by a physically motivated multi-view adversarial refinement stage to enhance spectral realism, envelope coherence, and harmonic consistency. Experiments on multiple datasets and bandwidth settings show that AnyBand consistently improves spectral reconstruction over existing baselines while achieving competitive perceptual quality across both standard and irregular input cutoffs. Audio samples are available\footnote{\url{https://danny-nus.github.io/AnyBand-DemoPage/}}.
\end{abstract}


\section{Introduction}
\label{sec:intro}

\begin{figure}[t]
    \centering
    \includegraphics[width=\linewidth]{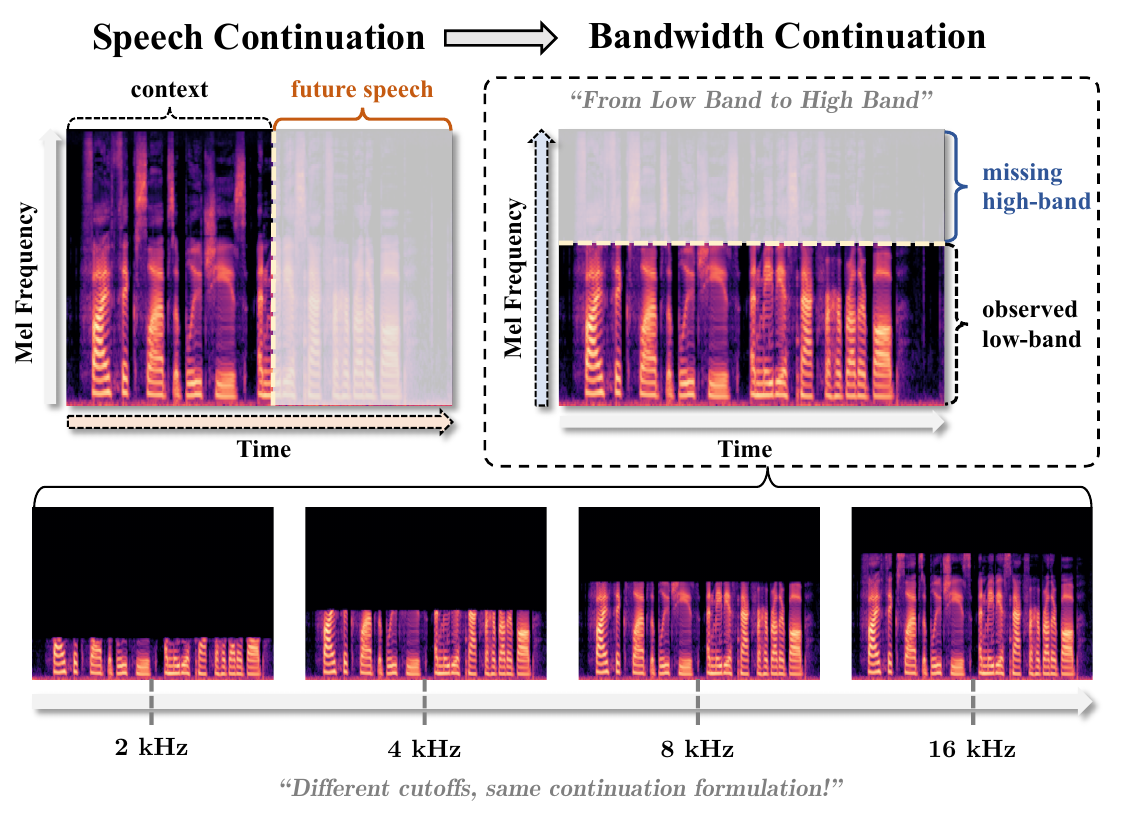}
    \caption{
        Bandwidth extension as frequency-domain continuation across different
        cutoff frequencies.
    }
    \label{fig:motivation}
\end{figure}

Speech bandwidth is often constrained by acquisition devices, communication channels, compression systems, or legacy recording conditions. Bandwidth extension (BWE) aims to restore missing high-frequency content from band-limited speech, thereby improving clarity, naturalness, and perceptual quality. The task is inherently underdetermined, since multiple acoustically plausible high-frequency continuations may correspond to the same observed low-frequency signal~\cite{yang2026survey}. Despite substantial advances in neural speech restoration and generative audio modeling, supporting diverse and continuously varying input cutoffs within a single BWE system remains challenging.

Many neural BWE and audio super-resolution methods have improved
high-frequency reconstruction under fixed or predefined bandwidth settings
using convolutional, WaveNet-based, flow-based, diffusion-based, adversarial,
or codec-based generative models
~\cite{kuleshov2017audio,gupta2019speech,li2021real,su2021bandwidth,
lee2021nu,zhang2021wsrglow,lu2024towards,fang2025vector,zhang2026codecflow,yang2026p2flow}.
Although these approaches achieve strong performance within their target
bandwidth configurations, their behavior is often closely coupled to the
source–target bandwidth configurations covered during training. Adapting to a different input cutoff may
therefore require retraining, fine-tuning, or additional model specialization,
limiting their flexibility in practical applications.

Recent works have explored unified, flexible, or blind BWE and audio
super-resolution systems that support multiple sampling rates, bandwidths,
degradation conditions, or unknown low-pass filters
~\cite{andreev2023hifi++,han2022nu,liu2022neural,liu2022voicefixer,
liu2024audiosr,salhab2025speech,sharma2026fast,moliner2024blind,kim2024audio}.
Although these methods reduce the need for cutoff-specific models, they
typically formulate varying bandwidths as separate restoration, upsampling, or
conditioning scenarios. We instead cast multi-bandwidth BWE as a unified
spectral-infilling problem, where the observed low-frequency spectrum provides
acoustic context and a frequency mask specifies the missing region. As shown
in Figure~\ref{fig:motivation}, different cutoffs correspond to different
amounts of observed context under the same frequency-continuation formulation.
The model then reconstructs the missing high-frequency content while preserving
the harmonic structure, voicing patterns, spectral envelope, and temporal
dynamics of the observed band.

We propose AnyBand, a unified framework that formulates BWE over continuously varying cutoff frequencies as in-context spectral infilling. Inspired by acoustic-context conditioning in prompt-based speech generation~\cite{wang2023neural,le2023voicebox,chen-etal-2025-f5,eskimez2024e2, zhao2026hierarchical, liang2026ted, zhao2026comelsinger}, AnyBand treats the observed low-frequency spectrum as a frequency-domain prompt and uses a cutoff-derived frequency mask to specify the missing band. We refer to this formulation as in-context spectral infilling because the observed spectrum provides acoustic context for generating the unobserved frequency region. A frequency-aware Diffusion Transformer models cross-frequency interactions and long-range temporal dependencies, while Easy-to-Balanced cutoff training and endpoint-focused adversarial refinement improve performance across cutoff conditions and enhance the realism of the generated high-frequency content.

Our contributions are summarized as follows:
\begin{itemize}
    \item We formulate multi-bandwidth speech extension as in-context spectral infilling, extending the temporal infilling of F5-TTS to the frequency axis: the observed low-frequency spectrum provides acoustic context, and a cutoff-derived mask defines the band to generate across continuously varying cutoff frequencies.

    \item We adapt the TTS backbone into a frequency-aware conditional flow that models cross-frequency and long-range temporal dependencies, together with an Easy-to-Balanced curriculum that uses the cutoff frequency as its difficulty variable.

    \item We introduce multi-view adversarial refinement with complementary objectives for spectral realism, cross-band envelope coherence, and harmonic consistency, applied through an efficient endpoint-focused rollout.
\end{itemize}
\section{Related Work}

\label{sec:related_work}

\subsection{Speech Bandwidth Extension and Super-Resolution}

Early neural BWE and audio super-resolution methods mainly learned direct
waveform- or spectrum-level mappings from band-limited to full-band speech. \cite{kuleshov2017audio} introduced a convolutional waveform
model, while TFNet~\cite{lim2018time} jointly modeled time- and
frequency-domain representations. Later work further improved efficiency for
real-time BWE~\cite{li2021real}.

Generative methods were subsequently introduced to improve high-frequency
reconstruction and perceptual quality, including autoregressive and
neural-vocoder-based models~\cite{gupta2019speech,su2021bandwidth},
normalizing flows~\cite{zhang2021wsrglow}, and diffusion models
~\cite{lee2021nu}. Other studies directly predict magnitude, phase, or complex
spectra~\cite{hu2020phase,mandel2023aero,lu2024towards}, often with
adversarial supervision~\cite{andreev2023hifi++,lu2024towards}. Recent work
also explores quantized or codec-derived latent spaces
~\cite{fang2025vector,zhang2026codecflow} and Schr\"odinger bridges
~\cite{kong2025a2sb}. However, many of these systems remain specialized for
fixed or predefined source--target bandwidth settings.

\subsection{Unified and Flexible Bandwidth Extension}

Recent studies have developed unified systems for multiple sampling rates,
bandwidths, or degradation conditions. NU-Wave 2~\cite{han2022nu} conditions
a diffusion model on input bandwidth, while MS-BWE~\cite{lu2024multi} uses
cascaded stages for flexible sampling-rate conversion. NVSR
~\cite{liu2022neural}, VoiceFixer~\cite{liu2022voicefixer}, and AudioSR
~\cite{liu2024audiosr} address broader super-resolution or restoration
settings. Other work improves flexibility or efficiency through adversarial
training, diffusion distillation, flow matching, or flexible vocoders
~\cite{salhab2025speech,im2025flashsr,choi2026universr,sharma2026fast, yun2025flowhigh}.
Blind BWE further handles unknown degradations; BABE
~\cite{moliner2024blind} jointly estimates a low-pass filter and reconstructs
full-band audio, but requires inference-time optimization.

A related direction uses partial acoustic context to guide speech generation.
Voicebox~\cite{le2023voicebox} infills missing temporal segments, while E2 TTS
~\cite{eskimez2024e2} and F5-TTS~\cite{chen-etal-2025-f5} use flow matching
conditioned on observed or reference speech. Inspired by this principle,
AnyBand treats the observed low-frequency spectrum as acoustic context and
uses an explicit frequency mask to formulate varying cutoffs as a common
spectral-infilling task. Closest to our formulation, Fre-Painter masks upper mel bands with varying
ratios to support multiple bandwidths, but relies on masked-autoencoder
pretraining, whereas AnyBand directly trains a missing-band conditional
flow conditioned on the observed band.
\section{Proposed Method}\label{sec:method}
\begin{figure*}[t]
    \centering
    \includegraphics[width=\textwidth]{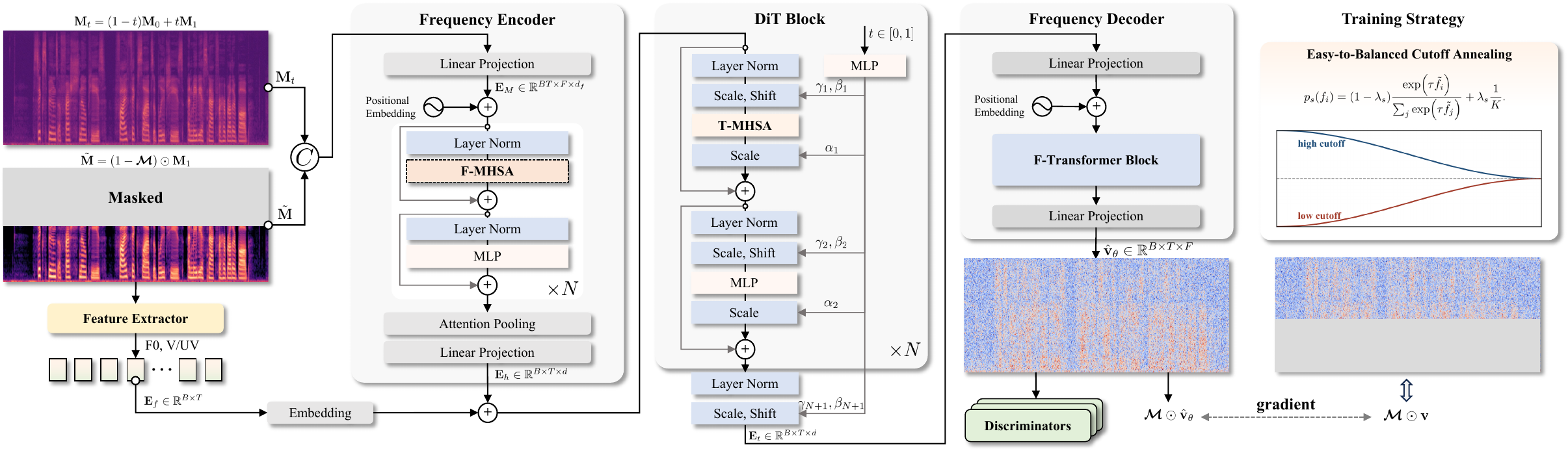}
    \caption{
        Overview of the proposed AnyBand framework.
        \textbf{AnyBand} encodes a noisy full-band spectral state together with
        a clean observed low-band prompt using a frequency encoder, models
        temporal dependencies with a DiT backbone, and predicts the
        mel-frequency velocity field using a frequency decoder. The model is
        trained with an Easy-to-Balanced cutoff curriculum and further refined
        using multi-view adversarial objectives.
    }
    \label{fig:overall}
\end{figure*}

\subsection{Problem Formulation: In-Context Spectral Infilling}
\label{sec:formulation}

Rather than treating each input cutoff as a separate source--target conversion
setting, AnyBand formulates multi-bandwidth BWE as \emph{in-context spectral
infilling}. The observed low-frequency spectrum serves as acoustic context,
while the masked high-frequency region specifies the region to be
reconstructed. Let
$\mathbf{M}\in\mathbb{R}^{F\times T}$ denote the target log-mel spectrogram,
where $F$ and $T$ are the numbers of mel bins and temporal frames. Given cutoff
frequency $f_c$ and its mel-bin index $k_c$, we define the missing-band mask
$\boldsymbol{\mathcal{M}}\in\{0,1\}^{F}$ as
$\mathcal{M}_k=\mathds{1}[k>k_c]$, where zeros and ones denote observed and
missing bins, respectively. The masked observation is:
\begin{equation}
\widetilde{\mathbf{M}}
=
\left(\mathbf{1}-\boldsymbol{\mathcal{M}}\right)\odot\mathbf{M},
\label{eq:masked_mel}
\end{equation}
with the mask broadcast over time. Thus, $\widetilde{\mathbf{M}}$ preserves
the spectrum below $f_c$ and removes the region to be reconstructed.

The retained spectrum provides cues about content, speaker identity, prosody,
and spectral structure, while its frequency extent specifies the BWE
condition. Conditioned on the clean spectral prompt
$\widetilde{\mathbf{M}}$ and auxiliary information $\mathbf{c}$, AnyBand
models
$p_{\theta}(\mathbf{M}\mid\widetilde{\mathbf{M}},\mathbf{c})$.
The frequency mask is used to construct the prompt and identify the
reconstruction region rather than being injected as a separate condition.
Training over continuously varying cutoffs enables a single model to handle
diverse bandwidth settings. At inference, the bandwidth is specified by the
extent of the observed spectrum without requiring a separate scalar cutoff
embedding.

\subsection{AnyBand}
\label{sec:anyband}

As illustrated in Figure~\ref{fig:overall}, AnyBand adopts the temporal DiT
backbone of F5-TTS~\cite{chen-etal-2025-f5}. Given an intermediate noisy
spectral state $\mathbf{M}_t$, the clean low-band prompt
$\widetilde{\mathbf{M}}$, a timestep $t$, and auxiliary conditioning
information $\mathbf{c}$, the generator directly predicts the velocity field
as:
\[
\hat{\mathbf{v}}_{\theta}
=
\mathcal{G}_{\theta}
\left(
\mathbf{M}_t,
\widetilde{\mathbf{M}},
t,
\mathbf{c}
\right).
\]
To explicitly capture dependencies along the frequency axis, we augment the
backbone with a frequency encoder and decoder placed before and after the
temporal DiT, respectively.

\paragraph{DiT Backbone.}
AnyBand retains the temporal DiT backbone of
F5-TTS~\cite{chen-etal-2025-f5}, including timestep conditioning and temporal
self-attention, while replacing its frame-wise mel projections with the
frequency encoder and decoder described below. Given the noisy spectral state
$\mathbf{M}_t$ and clean masked observation
$\widetilde{\mathbf{M}}$, the frequency encoder produces
$\mathbf{E}_h\in\mathbb{R}^{B\times T\times d}$, which is processed by the
DiT to model long-range temporal dependencies.

Following CodecFlow~\cite{zhang2026codecflow}, we further use F0 and
voiced/unvoiced information as auxiliary conditions for high-frequency
reconstruction. We define
$\mathbf{c}=[\mathbf{f}_0;\mathbf{v}]$, where
$\mathbf{f}_0\in\mathbb{R}^{B\times T}$ is the normalized F0 trajectory and
$\mathbf{v}\in\{0,1\}^{B\times T}$ is the voicing indicator. Their projected
embeddings are added to $\mathbf{E}_h$ before temporal modeling. The resulting
representations $\mathbf{E}_t\in\mathbb{R}^{B\times T\times d}$ are then
mapped back to the mel-frequency domain by the frequency decoder.

\paragraph{Frequency Encoder and Decoder.}
Bandwidth extension requires inferring missing high-frequency content from
frequency-dependent cues in the observed spectrum, including harmonic
structure, spectral envelopes, and cross-band continuity. AnyBand therefore
models the frequency axis explicitly before and after the temporal DiT
backbone. This layout follows the time--frequency factorization of speech
enhancement models~\cite{DBLP:conf/icassp/DangCZ22,DBLP:journals/taslp/WangCCLKW23}; we adapt it
to a flow-matching backbone, where frequency-axis attention relates
observed and missing bins within each frame.

For each frame, the frequency encoder treats the $F$ mel bins as frequency
tokens. Values from the noisy intermediate state $\mathbf{M}_t$ and clean
spectral prompt $\widetilde{\mathbf{M}}$ are concatenated, projected to
$d_f$-dimensional embeddings, and combined with learned frequency positional
embeddings, yielding
$\mathbf{E}_M^{(0)}
\in\mathbb{R}^{B\times T\times F\times d_f}$.
A shallow stack of pre-normalization Transformer blocks then updates the tokens
as:
\begin{equation}
\begin{aligned}
\mathbf{E}_M^{(l)\prime}
&=
\mathbf{E}_M^{(l-1)}
+
\operatorname{F\text{-}MHSA}
\left(
\operatorname{LN}
\left(
\mathbf{E}_M^{(l-1)}
\right)
\right), \\
\mathbf{E}_M^{(l)}
&=
\mathbf{E}_M^{(l)\prime}
+
\operatorname{FFN}
\left(
\operatorname{LN}
\left(
\mathbf{E}_M^{(l)\prime}
\right)
\right),
\end{aligned}
\label{eq:frequency_transformer}
\end{equation}
where $\operatorname{F\text{-}MHSA}$ applies self-attention along the
frequency axis independently at each frame. The final frequency tokens are
aggregated through attention pooling and projected to the DiT hidden
dimension, producing
$\mathbf{E}_h\in\mathbb{R}^{B\times T\times d}$.

After temporal modeling, the frequency decoder maps
$\mathbf{E}_t\in\mathbb{R}^{B\times T\times d}$ back to the mel-frequency
domain. Each frame representation is projected to the frequency hidden
dimension, broadcast over $F$ frequency positions, and combined with learned
frequency embeddings. The resulting tokens are processed by a frequency-axis
Transformer stack and mapped through a shared linear readout to produce
$\hat{\mathbf{v}}_{\theta}$. Thus, the encoder aggregates cross-frequency
evidence into frame-level representations, while the decoder predicts
frequency-specific velocities.

\begin{figure}[t]
    \centering
    \includegraphics[width=\linewidth]{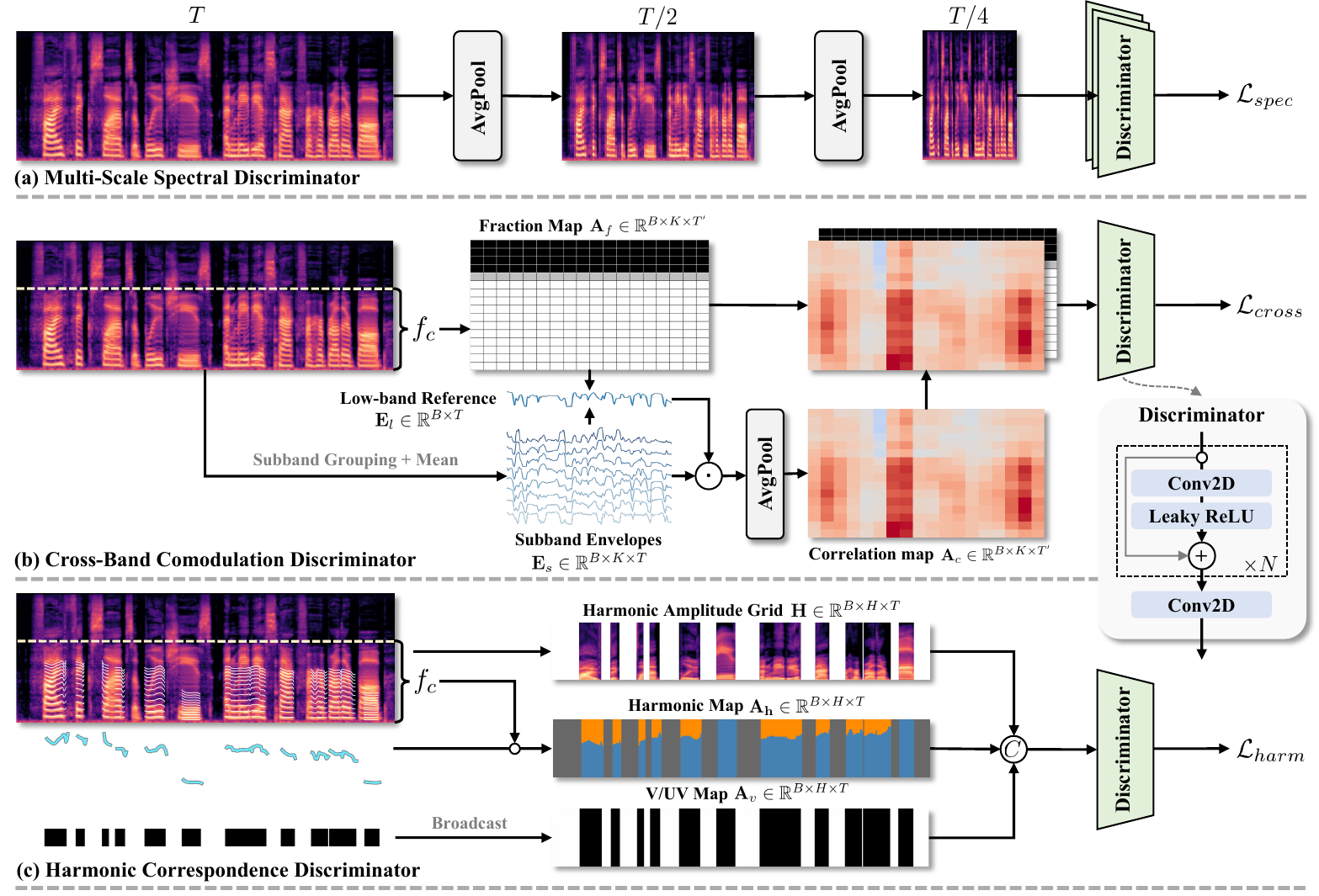}
    \caption{
        Overview of the proposed discriminators, which jointly assess
        multi-scale spectral realism, cross-band envelope dependencies, and
        $F_0$-aligned harmonic correspondence.
    }
    \label{fig:discriminator}
\end{figure}

\paragraph{Multi-View Spectral Discriminators.}
Although the masked flow-matching objective provides stable reconstruction
supervision, its pointwise loss may underemphasize perceptually important
high-frequency details. Inspired by adversarial training in neural speech
generation
~\cite{kong2020hifi,lee2022bigvgan,jang2021univnet,lee2024accelerating}, we
introduce three complementary discriminators:
$\mathcal{D}_{\mathrm{spec}}$,
$\mathcal{D}_{\mathrm{cross}}$, and
$\mathcal{D}_{\mathrm{harm}}$, targeting spectral realism, cross-band envelope
coherence, and harmonic correspondence, respectively.

The multi-scale spectral discriminator
$\mathcal{D}_{\mathrm{spec}}$ follows common multi-scale designs
~\cite{kong2020hifi,jang2021univnet}. Given a mel spectrogram
$\mathbf{M}$, we construct three inputs at temporal scales of
$1$, $1/2$, and $1/4$ by average pooling along time while preserving
frequency resolution. Independent branches then evaluate local
time--frequency patterns under different temporal receptive fields.

Motivated by cross-frequency envelope coherence in natural sounds
~\cite{nelken1999responses,mcdermott2011sound}, the cross-band discriminator
$\mathcal{D}_{\mathrm{cross}}$ evaluates whether reconstructed
high-frequency bands follow the temporal dynamics of the observed
low-frequency spectrum. We partition the $F$ mel bins into $K$ contiguous
subbands $\{\mathcal{B}_j\}_{j=1}^{K}$ and compute the envelope of the
$j$-th subband as:
\begin{equation}
\mathbf{E}_s^{(j)}
=
\frac{1}{|\mathcal{B}_j|}
\sum_{k\in\mathcal{B}_j}
\mathbf{M}_{k,:}.
\label{eq:subband_envelope}
\end{equation}
After temporal standardization, the subband envelopes form
$\mathbf{E}_s\in\mathbb{R}^{K\times T}$. We further compute the observed
fraction of each subband and the low-band reference envelope as:
\begin{equation}
\begin{aligned}
\rho_j
&=
\frac{1}{|\mathcal{B}_j|}
\sum_{k\in\mathcal{B}_j}
\left(1-\mathcal{M}_k\right), \\
\mathbf{E}_l
&=
\sum_{j=1}^{K}
\frac{\rho_j}
{\sum_{i=1}^{K}\rho_i+\epsilon}
\overline{\mathbf{E}}_s^{(j)},
\end{aligned}
\label{eq:low_band_reference}
\end{equation}
where $\overline{\mathbf{E}}_s^{(j)}$ denotes the standardized envelope.
Their element-wise products are locally averaged over time to form a
cross-band coherence map, which is combined with the observed-fraction map
and fed to $\mathcal{D}_{\mathrm{cross}}$.

Inspired by F0-guided source modeling
~\cite{lu2022source,li2023hiftnet,xu2025universal}, the harmonic discriminator
$\mathcal{D}_{\mathrm{harm}}$ evaluates whether reconstructed
high-frequency energy is consistent with the harmonic structure implied by
F0 and voicing. At frame $\tau$, the $n$-th harmonic frequency is
$f_n(\tau)=n f_0(\tau)$. Bilinear sampling along these trajectories yields an
$F_0$-aligned harmonic feature grid
$\mathbf{H}\in\mathbb{R}^{N_h\times T}$, which is stacked with an
observed-harmonic mask and a voiced/unvoiced map before being passed to
$\mathcal{D}_{\mathrm{harm}}$. All three discriminators are implemented with
lightweight 2D CNNs.

\subsection{Training and Sampling}
\label{sec:training_inference}

\paragraph{Easy-to-Balanced Cutoff Training.}
Uniform sampling from the beginning may expose the model to severely
band-limited inputs before it has learned basic spectral continuation. Following the curriculum-learning principle~\cite{DBLP:conf/icml/BengioLCW09},
we therefore introduce an Easy-to-Balanced curriculum that initially favors
higher cutoffs and gradually transitions to uniform sampling. Given the
normalized cutoff
$\widetilde{f}=(f_c-f_{\min})/(f_{\max}-f_{\min})\in[0,1]$, we sample:
\begin{equation}
\begin{aligned}
p_s(\widetilde{f})
&=
(1-\lambda_s)
\frac{\beta e^{\beta\widetilde{f}}}{e^\beta-1}
+\lambda_s,\\
\lambda_s
&=
\frac{1-\cos\!\left(
\pi\min\left(\frac{s}{\rho S},1\right)\right)}{2},
\end{aligned}
\label{eq:cutoff_curriculum}
\end{equation}
where $\beta$ controls the initial high-cutoff bias, $S$ is the total number
of training steps, and $\rho$ controls the annealing duration. The sampled
cutoff is mapped to the corresponding mel-bin boundary to construct the
frequency mask. As training progresses, the distribution smoothly approaches
uniform sampling over the full cutoff range.

\paragraph{Missing-Band Flow Training.}
Following conditional flow matching~\cite{lipman2022flow}, we sample
$t\in[0,1)$ and Gaussian noise
$\boldsymbol{\epsilon}\sim\mathcal{N}(\mathbf{0},\mathbf{I})$, and construct
the intermediate noisy state:
\begin{equation}
\mathbf{M}_t
=
t\mathbf{M}
+
(1-t)\boldsymbol{\epsilon}.
\label{eq:flow_path}
\end{equation}
The clean masked observation $\widetilde{\mathbf{M}}$ remains uncorrupted and
is provided separately as the spectral prompt. Under this linear probability
path, the target velocity is:
\begin{equation}
\mathbf{v}
=
\frac{\mathbf{M}-\mathbf{M}_t}{1-t}
=
\mathbf{M}-\boldsymbol{\epsilon}.
\label{eq:velocity_target}
\end{equation}
The generator directly predicts
$\hat{\mathbf{v}}_{\theta}
=
\mathcal{G}_{\theta}
(\mathbf{M}_t,\widetilde{\mathbf{M}},t,\mathbf{c})$.
Since the observed low-frequency region is already provided through
$\widetilde{\mathbf{M}}$, the training objective is applied only to the
missing frequency region:
\begin{equation}
\mathcal{L}_{\mathrm{flow}}
=
\mathbb{E}
\left[
\frac{
\left\|
\boldsymbol{\mathcal{M}}
\odot
\left(
\mathbf{v}
-
\hat{\mathbf{v}}_{\theta}
\right)
\right\|_2^2
}{
T \cdot \sum_{k=1}^{F}\mathcal{M}_k
}
\right],
\label{eq:flow_training_loss}
\end{equation}
where the expectation is taken over training samples, cutoff frequencies, flow
times, and Gaussian noise. The normalization by the number of missing
time--frequency elements keeps the loss scale comparable across different
cutoff frequencies.

\paragraph{Adversarial Refinement.}
After flow pretraining, we refine the generator with the proposed multi-view
spectral discriminators. Under the velocity-prediction parameterization, the
generator directly predicts $\hat{\mathbf{v}}_{\theta}$. At an intermediate
timestep $t$, the corresponding clean-spectrogram estimate is recovered as:
\begin{equation}
\hat{\mathbf{M}}
=
\mathbf{M}_t
+
(1-t)\hat{\mathbf{v}}_{\theta}.
\label{eq:clean_estimate}
\end{equation}
Rather than applying adversarial supervision to arbitrary intermediate states,
we construct fake samples from the model's own sampling trajectory. We first
integrate from Gaussian noise to an endpoint-near timestep $t_e$ without
gradient tracking, and then back-propagate through a short differentiable
rollout toward the data endpoint. The generated spectrogram is combined with
the observed band as: 
\begin{equation}
\hat{\mathbf{M}}_{\mathrm{fb}}
=
(1-\boldsymbol{\mathcal{M}})\odot\widetilde{\mathbf{M}}
+
\boldsymbol{\mathcal{M}}\odot\hat{\mathbf{M}},
\label{eq:fake_fullband_composition}
\end{equation}
so that the known region is copied from the input and only the missing band is
supplied by the model.

We apply adversarial refinement near the endpoint because earlier flow states
are noisy intermediate variables rather than natural spectrograms
~\cite{lipman2022flow}. This allows the discriminators to focus on spectral
realism and high-frequency detail without back-propagating through the full ODE
trajectory, following recent adversarial flow-matching audio generation
~\cite{lee2024accelerating}.

Let $\mathcal{Q}=\{\mathrm{spec},\mathrm{cross},\mathrm{harm}\}$ denote the
discriminator set. For each $\mathcal{D}_q$, we use a hinge GAN loss with
generator objective
$\mathcal{L}_{\mathrm{adv}}^{(q)}
=-\mathbb{E}[\mathcal{D}_q(\hat{\mathbf{M}}_{\mathrm{fb}})]$,
together with feature matching
$\mathcal{L}_{\mathrm{fm}}^{(q)}$. The generator is optimized with: 
\begin{equation}
\mathcal{L}_{G}
=
\mathcal{L}_{\mathrm{flow}}
+
\lambda_{\mathrm{adv}}
\sum_{q\in\mathcal{Q}}
\mathcal{L}_{\mathrm{adv}}^{(q)}
+
\lambda_{\mathrm{fm}}
\sum_{q\in\mathcal{Q}}
\mathcal{L}_{\mathrm{fm}}^{(q)} .
\label{eq:generator_total_loss}
\end{equation}

\paragraph{Sampling.}
At inference time, AnyBand initializes the full-band state with Gaussian noise
and solves the learned conditional flow over an inference time grid
$0=t_0<\cdots<t_N=1$, while the clean observed low-frequency spectrum is
provided separately as the spectral prompt. At the $i$-th integration step,
the generator directly predicts the conditional and unconditional velocity
fields, $\hat{\mathbf{v}}^{\mathrm{cond}}_i$ and
$\hat{\mathbf{v}}^{\emptyset}_i$, respectively. We apply classifier-free
guidance in the velocity space as: 
\begin{equation}
\hat{\mathbf{v}}^{\mathrm{cfg}}_i
=
\hat{\mathbf{v}}^{\emptyset}_i
+
w_{\mathrm{cfg}}
\left(
\hat{\mathbf{v}}^{\mathrm{cond}}_i
-
\hat{\mathbf{v}}^{\emptyset}_i
\right),
\label{eq:cfg_velocity}
\end{equation}
where $w_{\mathrm{cfg}}$ denotes the guidance scale. The guided velocity field
is integrated using the Heun solver. After the final integration step, the
observed low-frequency region is restored from the input prompt, while the
generated result is retained in the missing high-frequency region.

\section{Experiment Settings}
\begin{table*}[!t]
\centering
\caption{
Quantitative bandwidth-extension results on the in-domain VCTK and
out-of-domain EARS test sets.
All systems generate audio at 48~kHz.
The best and second-best results within each input setting are highlighted in
\textbf{bold} and \underline{underlined}, respectively.
}
\label{tab:main}

\setlength{\tabcolsep}{3.0pt}
\renewcommand{\arraystretch}{1.02}

\resizebox{\textwidth}{!}{
\begin{tabular}{clcccccc|cccccc}
\toprule
\multirow{2}{*}{\textbf{Input SR}}
& \multirow{2}{*}{\textbf{Method}}
& \multicolumn{6}{c|}{\textbf{VCTK}}
& \multicolumn{6}{c}{\textbf{EARS}} \\
\cmidrule(lr){3-8}
\cmidrule(lr){9-14}
&
& \textbf{LSD} $\downarrow$
& \textbf{LF-LSD} $\downarrow$
& \textbf{HF-LSD} $\downarrow$
& \textbf{NISQA} $\uparrow$
& \textbf{COL} $\uparrow$
& \textbf{STOI} $\uparrow$
& \textbf{LSD} $\downarrow$
& \textbf{LF-LSD} $\downarrow$
& \textbf{HF-LSD} $\downarrow$
& \textbf{NISQA} $\uparrow$
& \textbf{COL} $\uparrow$
& \textbf{STOI} $\uparrow$ \\
\midrule

\multirow{5}{*}{2 kHz}
& NU-Wave 2
& 2.685
& 1.364
& 2.727
& 1.897
& 2.115
& 0.7746
& 4.039
& 2.145
& 4.100
& 2.012
& 2.401
& 0.7285 \\

& AudioSR
& 2.373
& \underline{0.833}
& 2.417
& 2.356
& \underline{2.621}
& \underline{0.7913}
& \underline{1.958}
& \underline{1.408}
& \underline{1.971}
& 2.003
& 2.467
& 0.7444 \\

& FLowHigh
& 1.637
& 0.985
& 1.655
& 1.840
& 2.099
& 0.7885
& 2.304
& 1.775
& 2.319
& 1.622
& 1.565
& 0.7687 \\

& Fre-Painter
& \underline{1.323}
& 1.008
& \underline{1.331}
& \underline{2.697}
& 2.422
& 0.7820
& 2.568
& 1.730
& 2.595
& \underline{2.552}
& \underline{2.510}
& \underline{0.7748} \\

& AnyBand (Ours)
& \textbf{1.248}
& \textbf{0.539}
& \textbf{1.269}
& \textbf{3.125}
& \textbf{3.419}
& \textbf{0.8214}
& \textbf{1.546}
& \textbf{0.534}
& \textbf{1.574}
& \textbf{3.109}
& \textbf{2.973}
& \textbf{0.8067} \\
\midrule

\multirow{5}{*}{4 kHz}
& NU-Wave 2
& 2.552
& 1.258
& 2.636
& 2.806
& 3.168
& 0.8811
& 3.908
& 1.971
& 4.036
& 2.859
& 3.226
& 0.8372 \\

& AudioSR
& 1.628
& \underline{0.544}
& 1.691
& \underline{3.865}
& 3.700
& 0.8996
& \underline{1.291}
& \textbf{0.537}
& \underline{1.337}
& \textbf{3.739}
& \textbf{3.886}
& 0.8875 \\

& FLowHigh
& \underline{1.249}
& 0.808
& \underline{1.277}
& 3.784
& \underline{3.704}
& \underline{0.9291}
& 2.418
& 1.516
& 2.481
& 3.297
& 3.400
& 0.9069 \\

& Fre-Painter
& 1.286
& 0.873
& 1.312
& 3.539
& 3.408
& 0.9141
& 2.441
& 1.526
& 2.503
& 3.057
& 2.985
& \underline{0.9077} \\

& AnyBand (Ours)
& \textbf{1.180}
& \textbf{0.527}
& \textbf{1.219}
& \textbf{4.038}
& \textbf{3.966}
& \textbf{0.9356}
& \textbf{1.187}
& \underline{0.543}
& \textbf{1.228}
& \underline{3.690}
& \underline{3.681}
& \textbf{0.9205} \\
\midrule

\multirow{5}{*}{8 kHz}
& NU-Wave 2
& 2.391
& 1.248
& 2.551
& 3.100
& 3.397
& 0.9835
& 3.743
& 1.939
& 4.001
& 3.142
& 3.393
& 0.9433 \\

& AudioSR
& 1.528
& \underline{0.542}
& 1.654
& 3.937
& 3.732
& 0.9777
& \underline{1.173}
& \underline{0.558}
& \underline{1.256}
& 3.774
& 3.671
& 0.9713 \\

& FLowHigh
& 1.228
& 0.814
& 1.286
& \underline{4.002}
& \textbf{4.015}
& \textbf{0.9877}
& 2.337
& 1.492
& 2.464
& \underline{3.825}
& \textbf{3.769}
& \underline{0.9845} \\

& Fre-Painter
& \underline{1.208}
& 0.851
& \underline{1.258}
& 3.831
& 3.764
& 0.9861
& 2.351
& 1.478
& 2.478
& 3.715
& 3.670
& 0.9836 \\

& AnyBand (Ours)
& \textbf{1.086}
& \textbf{0.514}
& \textbf{1.155}
& \textbf{4.014}
& \underline{3.983}
& \underline{0.9870}
& \textbf{1.038}
& \textbf{0.526}
& \textbf{1.112}
& \textbf{3.834}
& \underline{3.737}
& \textbf{0.9847} \\
\midrule

\multirow{5}{*}{16 kHz}
& NU-Wave 2
& 2.065
& 1.287
& 2.344
& 3.716
& 3.578
& 0.9972
& 3.526
& 1.999
& 4.068
& 3.296
& 3.212
& 0.9553 \\

& AudioSR
& 1.415
& \underline{0.639}
& 1.644
& 3.828
& \textbf{3.921}
& 0.9967
& \underline{1.009}
& \underline{0.608}
& \underline{1.150}
& 3.799
& 3.698
& 0.9966 \\

& FLowHigh
& 1.140
& 0.833
& 1.246
& 3.723
& \underline{3.870}
& 0.9989
& 2.318
& 1.473
& 2.626
& 3.845
& \textbf{3.876}
& \underline{0.9990} \\

& Fre-Painter
& \underline{1.137}
& 0.909
& \underline{1.212}
& \underline{3.869}
& 3.772
& \textbf{0.9996}
& 2.299
& 1.490
& 2.592
& \underline{3.916}
& 3.836
& \textbf{0.9995} \\

& AnyBand (Ours)
& \textbf{0.974}
& \textbf{0.567}
& \textbf{1.092}
& \textbf{3.936}
& 3.837
& \underline{0.9992}
& \textbf{0.986}
& \textbf{0.583}
& \textbf{1.085}
& \textbf{3.922}
& \underline{3.875}
& 0.9964 \\

\bottomrule
\end{tabular}
}
\end{table*}
\begin{table}[t]
\centering
\caption{
Quantitative generalization results on VCTK with irregular input bandwidths.
Input sampling rates of 3, 6, 12, and 15\,kHz correspond to Nyquist
cutoffs of 1.5, 3, 6, and 7.5\,kHz, respectively.
The best and second-best results are highlighted in \textbf{bold} and
\underline{underlined}, respectively.
}
\label{tab:irregular}

\scriptsize
\setlength{\tabcolsep}{3.4pt}
\renewcommand{\arraystretch}{1.04}

\resizebox{\columnwidth}{!}{
\begin{tabular}{clccccc}
\toprule
\textbf{Input SR}
& \textbf{Method}
& \textbf{LSD} $\downarrow$
& \textbf{HF-LSD} $\downarrow$
& \textbf{NISQA} $\uparrow$
& \textbf{COL} $\uparrow$
& \textbf{STOI} $\uparrow$ \\
\midrule

\multirow{5}{*}{3\,kHz}
& NU-Wave 2
& 2.598
& 2.661
& 2.244
& 2.357
& 0.8291 \\

& AudioSR
& 1.643
& 1.690
& \underline{3.318}
& \underline{3.118}
& 0.8495 \\

& FLowHigh
& 1.548
& 1.575
& 2.394
& 2.302
& 0.8412 \\

& Fre-Painter
& \underline{1.298}
& \underline{1.315}
& 2.995
& 2.687
& \underline{0.8502} \\

& AnyBand (Ours)
& \textbf{1.228}
& \textbf{1.261}
& \textbf{3.671}
& \textbf{3.702}
& \textbf{0.8523} \\
\midrule

\multirow{5}{*}{6\,kHz}
& NU-Wave 2
& 2.462
& 2.586
& 3.172
& 3.497
& 0.9442 \\

& AudioSR
& 1.585
& 1.680
& 3.915
& 3.818
& 0.9500 \\

& FLowHigh
& \underline{1.234}
& \underline{1.278}
& \underline{3.948}
& \underline{3.891}
& 0.9502 \\

& Fre-Painter
& 1.259
& 1.302
& 3.654
& 3.546
& \underline{0.9541} \\

& AnyBand (Ours)
& \textbf{1.123}
& \textbf{1.177}
& \textbf{4.060}
& \textbf{4.002}
& \textbf{0.9575} \\
\midrule

\multirow{5}{*}{12\,kHz}
& NU-Wave 2
& 2.207
& 2.428
& 3.219
& 2.803
& 0.9903 \\

& AudioSR
& 1.458
& 1.635
& 3.878
& 3.801
& 0.9955 \\

& FLowHigh
& \underline{1.183}
& 1.263
& \textbf{3.996}
& \textbf{3.944}
& \underline{0.9963} \\

& Fre-Painter
& 1.198
& \underline{1.255}
& 3.849
& 3.785
& 0.9952 \\

& AnyBand (Ours)
& \textbf{1.002}
& \textbf{1.093}
& \underline{3.987}
& \underline{3.920}
& \textbf{0.9968} \\
\midrule

\multirow{5}{*}{15\,kHz}
& NU-Wave 2
& 2.094
& 2.360
& 3.558
& 3.396
& 0.9912 \\

& AudioSR
& 1.415
& 1.644
& 3.866
& 3.725
& 0.9967 \\

& FLowHigh
& \underline{1.151}
& 1.251
& 3.710
& 3.656
& \underline{0.9990} \\

& Fre-Painter
& 1.167
& \underline{1.242}
& \underline{3.954}
& \underline{3.868}
& \textbf{0.9993} \\

& AnyBand (Ours)
& \textbf{0.975}
& \textbf{1.085}
& \textbf{3.975}
& \textbf{3.894}
& 0.9987 \\

\bottomrule
\end{tabular}
}
\end{table}

\subsection{Datasets}
\label{subsec:data}

We conduct experiments on the 48-kHz VCTK~\cite{yamagishi2019vctk} and
EARS~\cite{richter2024ears} datasets. Following CodecFlow~\cite{zhang2026codecflow},
VCTK speakers p225--p228 are held out for in-domain evaluation of all
methods, while the remaining speakers are used for training.
For out-of-domain evaluation, we use 50 utterances from 10 EARS speakers.
Following CodecFlow and SpecMaskGIT~\cite{DBLP:conf/ismir/ComunitaZTYZSI024}, band-limited inputs
are constructed by masking mel bins above a given cutoff; this setting
targets cutoff variation rather than codec or channel distortions. F0 and
V/UV are extracted from the band-limited waveform obtained by low-pass
filtering and resampling, so the pitch features never use the clean target.
Cutoffs are sampled continuously from 1 to 16~kHz during training and
evaluated under both standard and irregular settings; the irregular input
rates (3, 6, 12, and 15~kHz) are nonstandard rates within this range rather
than unseen cutoffs. The reported input sampling rate equals twice the
cutoff frequency.

\subsection{Implementation Details}
\label{subsec:imp}

We use 128-bin mel spectrograms computed at 48~kHz with a 2,048-point FFT and
a hop length of 256. AnyBand contains 8 temporal DiT blocks with hidden
dimension 384 and 6 attention heads. Its frequency encoder and decoder each
use 2 frequency-axis attention blocks with hidden dimension 64 and 4 heads.
Following PACE~\cite{zhao2025prosody}, frame-level F0 and voiced/unvoiced
features extracted from the input speech are used as auxiliary conditions.
All discriminators are lightweight 2D CNNs with Leaky-ReLU activations;
branches with the same architecture use independent parameters. Training uses
AdamW on eight NVIDIA L40S GPUs with a batch size of 32 per GPU. We train the
flow model for 300 epochs with the Easy-to-Balanced cutoff curriculum, using
$\rho=0.7$ and $\beta=4$, followed by 200 epochs of adversarial refinement
with uniformly sampled continuous cutoffs. At inference, we use the Heun
solver with 50 steps and a classifier-free guidance scale of 1.4. Generated
mel spectrograms are converted to 48-kHz waveforms using
Vocos\footnote{\url{https://huggingface.co/kittn/vocos-mel-48khz-alpha1}}
~\cite{siuzdak2023vocos}.

\subsection{Evaluation Metrics}
\label{subsec:metrics}
We evaluate spectral reconstruction using log-spectral distance
(LSD)~\cite{gray1976distance} and its low- and high-frequency variants
~\cite{han2022nu}. LSD is computed over the full spectrum, while LF-LSD and
HF-LSD measure distortion in the observed and missing frequency regions,
respectively. Lower values indicate better reconstruction. We further report
NISQA and its coloration dimension (COL)~\cite{mittag2021nisqa} for overall
perceptual quality and frequency-response distortion, and
STOI~\cite{taal2011algorithm} for intelligibility preservation. For subjective
evaluation, 15 listeners rated the sound quality of outputs from each system
and the ground truth on a five-point scale, using five utterances at each
input sampling rate of 8, 12, 16, 20, and 24~kHz. 

\section{Results}
\subsection{Main Results}
\label{sec:main_results}

\paragraph{Objective Results.} Table~\ref{tab:main} compares AnyBand with representative bandwidth-extension
systems \cite{han2022nu, liu2024audiosr, kim2024audio, yun2025flowhigh} on the in-domain VCTK and out-of-domain EARS test sets. AnyBand
achieves the lowest LSD and HF-LSD across all tested cutoff settings,
demonstrating consistently strong full-band and high-frequency reconstruction.
Its low LF-LSD further suggests that the observed low-frequency content is
well preserved. The spectral gains remain consistent on EARS, indicating good
cross-domain generalization, and are most pronounced under severely
band-limited conditions. At higher input bandwidths, the differences in
perceptual metrics become smaller, but AnyBand remains competitive in NISQA,
COL, and STOI while retaining a clear spectral advantage.

Table~\ref{tab:irregular} further evaluates performance at irregular cutoff
settings within the continuous training range. AnyBand achieves the best LSD
and HF-LSD at every tested cutoff, confirming that a single model can support
nonstandard bandwidth configurations without cutoff-specific specialization.
The improvements are again larger at lower cutoffs, where a greater portion of
the spectrum must be reconstructed, while perceptual quality and
intelligibility remain competitive. Overall, these results support
mask-specified spectral infilling as an effective formulation for flexible BWE
across both standard and irregular cutoff settings.

\begin{figure}[t]
    \centering
    \includegraphics[width=0.9\columnwidth]
    {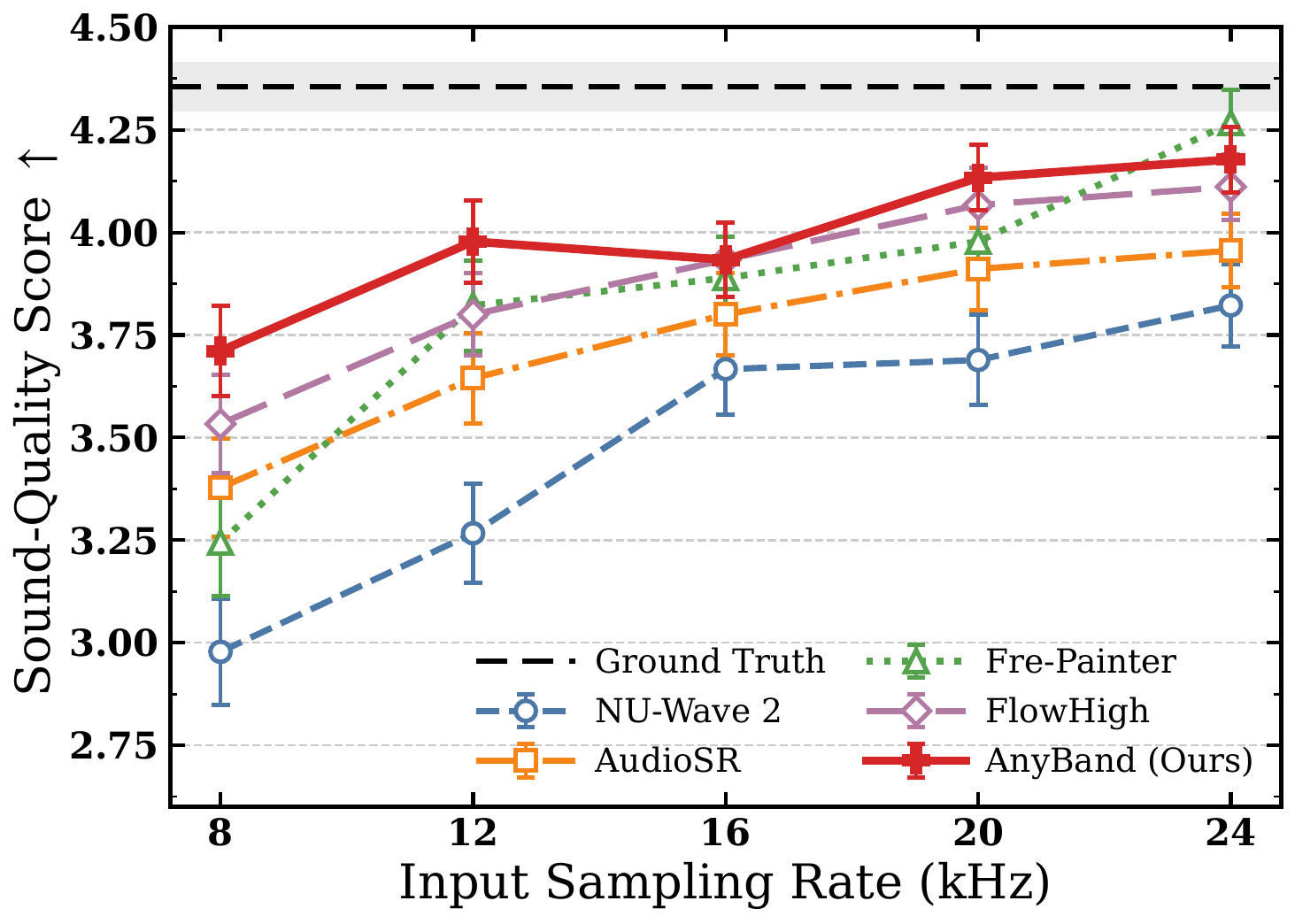}
    \caption{
    Subjective scores across different input sampling rates.
    The dashed horizontal line denotes the ground-truth score, and error bars
    indicate 95\% confidence intervals.
    }
    \label{fig:subjective_quality}
\end{figure}

\paragraph{Subjective Evaluation.} Figure~\ref{fig:subjective_quality} shows that AnyBand achieves the highest
subjective quality at most input sampling rates and remains competitive across
all evaluated bandwidths. Its advantage is more evident for severely
band-limited inputs, indicating stronger perceptual reconstruction when a
larger portion of the spectrum is missing. At higher input sampling rates, the
differences among methods become smaller, while all systems remain below the
ground-truth quality.

\begin{table}[t]
\centering
\caption{
Ablation study of cutoff-sampling strategies on VCTK. Input sampling rates
of 2, 4, 8, and 16~kHz correspond to Nyquist cutoffs of 1, 2, 4, and 8~kHz,
respectively. The output sampling rate is 48~kHz. The best and second-best
results within each input setting are highlighted in \textbf{bold} and
\underline{underlined}, respectively.
}
\label{tab:curriculum}

\scriptsize
\setlength{\tabcolsep}{2.4pt}
\renewcommand{\arraystretch}{1.02}

\resizebox{\columnwidth}{!}{
\begin{tabular}{clccccc}
\toprule
\textbf{Input SR}
& \textbf{Sampling strategy}
& \textbf{LSD} $\downarrow$
& \textbf{HF-LSD} $\downarrow$
& \textbf{NISQA} $\uparrow$
& \textbf{COL} $\uparrow$
& \textbf{STOI} $\uparrow$ \\
\midrule

\multirow{3}{*}{2 kHz}
& Uniform (discrete)
& \underline{1.284}
& \underline{1.302}
& \textbf{3.165}
& \underline{3.232}
& \textbf{0.8224} \\

& Uniform (continuous)
& 1.358
& 1.386
& 2.807
& 3.036
& 0.8052 \\

& Easy-to-balanced
& \textbf{1.248}
& \textbf{1.269}
& \underline{3.125}
& \textbf{3.419}
& \underline{0.8214} \\
\midrule

\multirow{3}{*}{4 kHz}
& Uniform (discrete)
& \underline{1.216}
& \underline{1.260}
& \underline{3.855}
& \underline{3.659}
& \underline{0.9287} \\

& Uniform (continuous)
& 1.248
& 1.295
& 3.698
& 3.567
& 0.9185 \\

& Easy-to-balanced
& \textbf{1.180}
& \textbf{1.219}
& \textbf{4.038}
& \textbf{3.966}
& \textbf{0.9356} \\
\midrule

\multirow{3}{*}{8 kHz}
& Uniform (discrete)
& \underline{1.101}
& \textbf{1.143}
& \underline{3.907}
& \underline{3.839}
& \underline{0.9861} \\

& Uniform (continuous)
& 1.132
& 1.199
& 3.826
& 3.799
& 0.9842 \\

& Easy-to-balanced
& \textbf{1.086}
& \underline{1.155}
& \textbf{4.014}
& \textbf{3.983}
& \textbf{0.9870} \\
\midrule

\multirow{3}{*}{16 kHz}
& Uniform (discrete)
& \textbf{0.970}
& \textbf{1.080}
& \textbf{4.011}
& \textbf{3.926}
& \textbf{0.9993} \\

& Uniform (continuous)
& 0.993
& 1.094
& 3.847
& 3.706
& 0.9980 \\

& Easy-to-Balanced
& \underline{0.974}
& \underline{1.092}
& \underline{3.936}
& \underline{3.837}
& \underline{0.9992} \\
\midrule

\multirow{3}{*}{Avg.}
& Uniform (discrete)
& \underline{1.143}
& \underline{1.196}
& \underline{3.735}
& \underline{3.664}
& \underline{0.9341} \\

& Uniform (continuous)
& 1.183
& 1.244
& 3.545
& 3.527
& 0.9265 \\

& Easy-to-balanced
& \textbf{1.122}
& \textbf{1.184}
& \textbf{3.778}
& \textbf{3.801}
& \textbf{0.9358} \\

\bottomrule
\end{tabular}
}
\end{table}

\begin{table}[t]
\centering
\caption{
Ablation study of the proposed components on VCTK with a 16-kHz input
sampling rate (8-kHz Nyquist cutoff). The best and second-best results are
highlighted in \textbf{bold} and \underline{underline}, respectively.
}
\label{tab:ablation}
\scriptsize
\setlength{\tabcolsep}{2.6pt}
\renewcommand{\arraystretch}{1.05}
\begin{tabular}{lccccc}
\toprule
\textbf{Variants}
& \textbf{LSD} $\downarrow$
& \textbf{HF-LSD} $\downarrow$
& \textbf{NISQA} $\uparrow$
& \textbf{COL} $\uparrow$
& \textbf{STOI} $\uparrow$ \\
\midrule

AnyBand (Ours)
& \textbf{0.974}
& \underline{1.092}
& \textbf{3.936}
& \textbf{3.837}
& \underline{0.9992} \\

\quad $-$ w/o Freq. Modules
& 1.038
& 1.159
& 3.772
& 3.656
& 0.9978 \\

\quad $-$ w/o F0/UV
& 1.035
& 1.125
& 3.758
& 3.698
& 0.9985 \\

\quad $-$ w/o $\mathcal{D}_{\mathrm{spec}}$
& 1.124
& 1.206
& 3.673
& 3.628
& 0.9935 \\

\quad $-$ w/o $\mathcal{D}_{\mathrm{cross}}$
& \underline{0.998}
& \textbf{1.076}
& \underline{3.847}
& \underline{3.752}
& 0.9968 \\

\quad $-$ w/o $\mathcal{D}_{\mathrm{harm}}$
& 1.012
& 1.098
& 3.760
& 3.735
& \textbf{0.9996} \\

\bottomrule
\end{tabular}
\end{table}

\subsection{Ablation Studies}
\label{sec:ablation}

To better understand the main design choices in AnyBand, we analyze the
cutoff-sampling strategy, model components, prediction parameterization, and
sampling configuration.

\paragraph{Cutoff-sampling strategy.}
Table~\ref{tab:curriculum} compares three training strategies. Discrete
uniform sampling draws equivalent input rates from
$\{2,4,8,16,24,32\}$~kHz, whereas continuous uniform sampling covers the full
cutoff range. Easy-to-Balanced starts from a distribution biased toward
higher-cutoff, less underdetermined examples and gradually transitions to
continuous uniform sampling. Discrete uniform remains competitive at its
predefined training points, while continuous uniform produces weaker average
results. Easy-to-Balanced achieves the best overall averages, with its largest
gains at low and medium cutoff settings, indicating a better balance across
bandwidth conditions.

\paragraph{Component analysis.}
Table~\ref{tab:ablation} evaluates the main components of AnyBand. In the
variant without frequency modules, the frequency encoder and decoder are
replaced by the linear input and output projections used in
F5-TTS~\cite{chen-etal-2025-f5}. The remaining variants remove F0/UV
conditioning or one of the three discriminators. Removing the frequency
modules or F0/UV conditioning generally degrades both spectral and perceptual
metrics. Among the discriminators, removing
$\mathcal{D}_{\mathrm{spec}}$ causes the largest overall degradation.
Although removing $\mathcal{D}_{\mathrm{cross}}$ or
$\mathcal{D}_{\mathrm{harm}}$ yields isolated improvements on individual
metrics, each removal worsens the overall metric profile, suggesting that the
three discriminators provide complementary supervision.


\begin{figure}[!t]
    \centering
    \begin{subfigure}[t]{0.49\columnwidth}
        \centering
        \includegraphics[width=\linewidth]{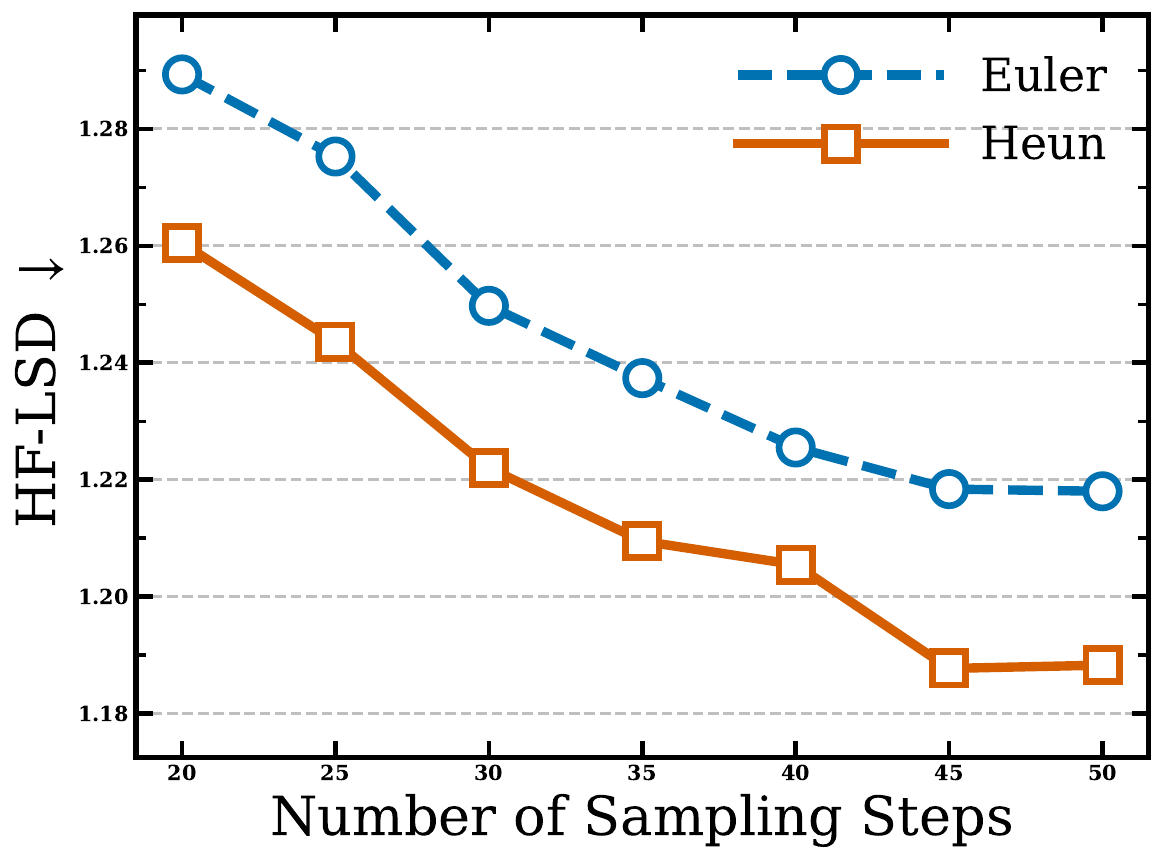}
        \caption{Solver and steps.}
        \label{fig:sampling_steps}
    \end{subfigure}
    \hfill
    \begin{subfigure}[t]{0.49\columnwidth}
        \centering
        \includegraphics[width=\linewidth]{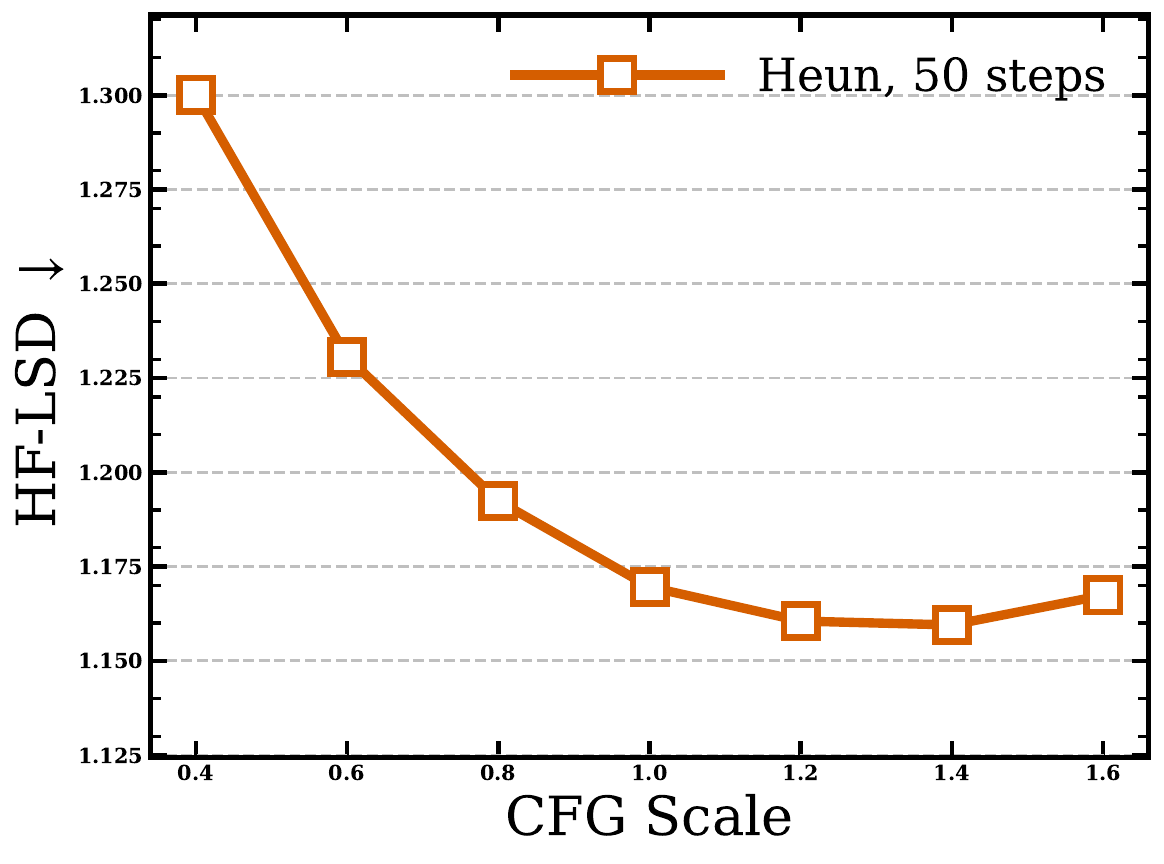}
        \caption{Guidance scale.}
        \label{fig:cfg_scale}
    \end{subfigure}
    \caption{Effect of sampling configurations on HF-LSD.}
    \label{fig:sampling_analysis}
\end{figure}

\paragraph{Sampling configuration.}
Figure~\ref{fig:sampling_analysis} examines the sampling solver, number of
steps, and classifier-free guidance scale. Increasing the number of steps
consistently reduces HF-LSD for both Euler and Heun, with Heun performing
better across the tested settings. The improvements diminish as the number of
steps increases, revealing a trade-off between reconstruction quality and
inference cost. Increasing the guidance scale initially improves
high-frequency reconstruction, but the benefit saturates and slightly
degrades at larger values. We therefore use the Heun solver with 50 steps and
a guidance scale of 1.4. 

\section{Conclusion}
We introduced \textbf{AnyBand}, a unified framework for speech bandwidth
extension over continuously varying cutoff frequencies. By formulating BWE as
in-context spectral infilling, AnyBand uses the observed low-frequency spectrum
as an acoustic prompt and a cutoff-derived mask to specify the missing region,
enabling a single model to handle diverse bandwidth settings. Its
frequency-aware encoder--decoder, F0/UV conditioning, and complementary
spectral, cross-band, and harmonic discriminators improve spectral
reconstruction and high-frequency realism. We further introduce an
Easy-to-Balanced cutoff curriculum and adopt velocity prediction for effective
flow-based generation. Experiments on in-domain and out-of-domain datasets show
that AnyBand consistently reduces full-band and high-frequency distortion over
prior systems while maintaining competitive perceptual quality and
intelligibility. These results support unified spectral infilling as an
effective formulation for flexible multi-bandwidth speech extension. 

\bibliography{aaai2027}

\appendix
\setcounter{secnumdepth}{2}
\section{Additional Implementation Details}
\label{sec:appendix_implementation}
\subsection{Data and Acoustic Features}

All speech signals are processed at 48~kHz. During training, we randomly
extract waveform segments of 64,000 samples. We follow the acoustic preprocessing
configuration of the 48-kHz Vocos model and compute 128-bin mel spectrograms
using a 2,048-point FFT and a hop length of 256. The mel filterbank covers
0--24~kHz and uses a magnitude power of 1.0, Slaney normalization, and the
Slaney mel scale.

The cutoff frequency is independently sampled for each training segment from
the continuous range of 1--16~kHz. The corresponding mel-bin boundary is
determined by comparing the center frequency of each mel bin with the sampled
cutoff, following the mask definition in the main paper.

Frame-level F0 and voiced/unvoiced features are extracted from the
corresponding band-limited input waveform using \texttt{pYIN}, with
\texttt{YIN} used as a fallback when \texttt{pYIN} does not return a valid
estimate. For voiced frames, F0 is logarithmically normalized as: 
\begin{equation}
    \widetilde f_0
    =
    \operatorname{clip}
    \left(
        \frac{\log f_0-\log 86}
             {\log 1100-\log 86},
        0,1
    \right).
\end{equation}
The normalized F0 value is set to zero for unvoiced frames, while a separate
binary feature indicates the voiced/unvoiced state.

\subsection{Model Configuration}

\paragraph{Generator.}
The AnyBand generator contains 22.25M parameters, including the frequency
encoder and decoder, which add only 0.09M parameters relative to the
frame-wise linear projections they replace. Waveforms are synthesized by
a 48-kHz Vocos vocoder (36.70M parameters), so inference uses 58.95M
parameters in total. Its
temporal backbone consists of eight DiT blocks with a hidden dimension of
384 and six attention heads. The frequency encoder and decoder each contain
two frequency-axis Transformer blocks with a hidden dimension of 64 and four
attention heads.
The frequency encoder takes the intermediate noisy state
$\mathbf{M}_t$ and the observed spectral prompt
$\widetilde{\mathbf{M}}$ as input. Their values are concatenated at each
time--frequency position and projected into frequency-token embeddings.
After frequency-axis self-attention, a learned query performs four-head
attention pooling over the frequency tokens to obtain a frame-level
representation. The temporal DiT then models long-range dependencies across
frames, after which the frequency decoder produces frequency-specific
velocity predictions. The frequency mask is used to construct the observed
prompt and define the reconstruction region, but is not directly injected
into the generator.

\paragraph{Discriminators.}
The multi-scale spectral discriminator $\mathcal{D}_{\mathrm{spec}}$ evaluates mel
spectrograms at three temporal resolutions. Each resolution is processed by
an independent 2D convolutional branch with channel dimensions of
32, 64, 128, and 256.
The cross-band discriminator $\mathcal{D}_{\mathrm{cross}}$ partitions the 128 mel
bins into 16 contiguous subbands, each containing eight mel bins, and
computes local cross-band statistics using a temporal window of 32 frames.
The harmonic discriminator $\mathcal{D}_{\mathrm{harm}}$ considers the first 12
F0-aligned harmonics, whose amplitudes are bilinearly sampled from the mel
spectrogram. The three discriminators use
separate, task-specific 2D convolutional networks.

\subsection{Training Configuration}

\paragraph{Flow Pretraining.}
We first train the generator for 300 epochs using the missing-band
velocity-prediction objective described in the main paper. The model is
optimized with AdamW using an initial learning rate of $10^{-3}$ and a
cosine-annealing learning-rate schedule with $T_{\max}=300$. No
learning-rate warmup is used. The Easy-to-Balanced cutoff curriculum uses
$\beta=4$ and $\rho=0.7$. During Stage~1, the observed-band prompt and the F0/VUV condition are
dropped jointly with probability 0.1 to train the unconditional branch used
for classifier-free guidance.

\paragraph{Adversarial Refinement.}
All discriminators use hinge losses; $\mathcal{D}_{\mathrm{harm}}$, conditioned
on F0/VUV $\mathbf{c}$, also treats real spectra paired with in-batch shuffled
$\tilde{\mathbf{c}}$ as negatives:
\begin{equation}
\begin{aligned}
\mathcal{L}_{D}={}&\textstyle\sum_{q\in\mathcal{Q}}\mathbb{E}\big[\phi(1-\mathcal{D}_q(\mathbf{M}))+\phi(1+\mathcal{D}_q(\hat{\mathbf{M}}_{\mathrm{fb}}))\big]\\
&+\mathbb{E}\big[\phi(1+\mathcal{D}_{\mathrm{harm}}(\mathbf{M},\tilde{\mathbf{c}}))\big],
\end{aligned}
\end{equation}
where $\phi(z)=\max(0,z)$.
The pretrained generator is subsequently refined for 200 epochs using the
three proposed discriminators. During this stage, cutoff frequencies are
uniformly sampled from the continuous training range. The generator and
discriminators are optimized with AdamW using learning rates of
$5\times10^{-5}$ and $2\times10^{-4}$, respectively, with
$(\beta_1,\beta_2)=(0.8,0.99)$. Both optimizers use cosine-annealing
schedules with $T_{\max}=200$ and no learning-rate warmup. The generator
and discriminators are updated once per batch in a 1:1 ratio.
We set the adversarial-loss weight to
$\lambda_{\mathrm{adv}}=0.1$ and equally weight the adversarial losses from
the three discriminators. The feature-matching weight is set to
$\lambda_{\mathrm{fm}}=2.0$, with feature-matching losses summed over their
intermediate feature maps.

For endpoint-focused adversarial refinement, we first integrate the flow
from Gaussian noise to $t_e=0.8$ without gradient tracking. Starting from
the resulting intermediate state, we perform an eight-step differentiable
Heun rollout over $[t_e,1]$. The adversarial and feature-matching objectives
are applied to the resulting full-band estimate, with the observed frequency
region copied from the band-limited input as defined in the main paper.

\paragraph{Computational Setup.}
AnyBand is trained with bfloat16 mixed precision on eight NVIDIA L40S GPUs,
using a batch size of 32 per GPU and a global batch size of 256.

\subsection{Inference Configuration}

At inference time, we use the Heun solver with 50 sampling steps and a
right-swayed time grid with $\gamma=2.0$. Classifier-free guidance is applied
in velocity space with a guidance scale of 1.4. The observed spectral prompt
is provided separately throughout sampling.
The observed frequency region is not overwritten after each intermediate
integration step and is copied back only after the final step. We generate
one output for each utterance in both objective and subjective evaluations. RTF is measured end-to-end, including F0/VUV extraction, 50-step Heun sampling
with classifier-free guidance, and Vocos synthesis. Excluding F0/VUV
extraction, the RTF of AnyBand is 0.4824. Baselines are timed with their
official inference code under the same setting.

\section{Experimental Details}
\label{sec:appendix_exp}
\subsection{Baseline Configurations}
\label{sec:appendix_baselines}

\paragraph{Common Protocol.}
All baseline systems are implemented using their official codebases. We use
the same VCTK training split, training utterances, held-out test utterances,
and nominal cutoff settings for all methods. To preserve the original
formulation of each baseline, band-limited inputs are constructed using the
native preprocessing procedure provided by its official implementation.
Methods initialized from publicly released checkpoints are subsequently
fine-tuned on the shared VCTK training split, whereas the remaining methods
are trained from scratch on the same data. All systems generate 48-kHz
outputs and are evaluated using the same metric protocols.

\paragraph{NU-Wave 2 \cite{han2022nu}.}
We use the official implementation of NU-Wave 2%
\footnote{\url{https://github.com/maum-ai/nuwave2}}
and train the model from scratch on the shared VCTK training split. We
retain its original bandwidth-conditioning mechanism and native waveform
preprocessing procedure, while adapting the training and evaluation cutoff
settings to our experimental protocol.

\paragraph{AudioSR \cite{liu2024audiosr}.}
We use the official AudioSR implementation and its publicly released
checkpoint%
\footnote{\url{https://github.com/haoheliu/versatile_audio_super_resolution}},
followed by fine-tuning on the shared VCTK training split. We retain the
official input preprocessing and recommended inference configuration.

\paragraph{FLowHigh \cite{yun2025flowhigh}.}
We use the official FLowHigh implementation%
\footnote{\url{https://github.com/resemble-ai/flowhigh}}
and initialize the model from the released checkpoint%
\footnote{\url{https://drive.google.com/drive/folders/1clsJ3bFTZSCLOb4I0wk0l6FL9Yh33Vof}}.
The model is then fine-tuned on the shared VCTK training split while
retaining its original probability path, input preprocessing, and inference
procedure.

\paragraph{Fre-Painter \cite{kim2024audio}.}
We use the official implementation of Fre-Painter%
\footnote{\url{https://github.com/ishine/FrePainter}}
and train the model from scratch on the shared VCTK training split. We
retain its original preprocessing, training objective, and inference
procedure, while adapting the training and evaluation bandwidth settings to
our experimental protocol.

\subsection{Ablation Configurations}
\label{sec:appendix_ablation_settings}

\paragraph{Component Ablations.}
For the variant without frequency modules, we replace the proposed
frequency encoder and decoder with the frame-wise linear input and output
projections used in the original temporal DiT backbone. 
The parameter-matched control augments the w/o-Freq.\ variant with four
frame-wise residual MLP blocks, reaching 22.2542M generator parameters
versus 22.2540M for the full model, and follows the same training and
evaluation protocol. 
The variant without
F0/VUV removes both auxiliary conditioning features while retaining the
remaining generator architecture. To evaluate the contribution of each adversarial objective, we remove one
discriminator at a time during adversarial refinement while retaining the
other two discriminators. All discriminator ablations are initialized from
the same Stage-1 flow checkpoint and independently refined for the same training steps. The remaining training data, cutoff-sampling
strategy, optimization settings, and inference configuration are kept
identical to those of the full model.

\paragraph{Cutoff-Sampling Strategies.}
We compare three cutoff-sampling strategies during Stage-1 flow
pretraining. \emph{Discrete Uniform} uniformly samples equivalent input
rates from $\{2,4,8,16,24,32\}$~kHz, corresponding to cutoff frequencies
of $\{1,2,4,8,12,16\}$~kHz. \emph{Continuous Uniform} samples cutoff
frequencies uniformly from the full continuous range of 1--16~kHz
throughout training. \emph{Easy-to-Balanced} initially favors higher
cutoff frequencies and gradually transitions to the same continuous uniform
distribution.

Continuous Uniform and Easy-to-Balanced share the same continuous cutoff
support and differ only in how the sampling distribution evolves during
Stage-1 training. All three variants use the same model architecture,
training data, optimization settings, and training budget. They are then
refined using the same Stage-2 adversarial-training configuration, in which
cutoff frequencies are uniformly sampled from the continuous training range.
Therefore, the compared models differ only in the cutoff-sampling strategy
used during Stage-1 flow pretraining.

\subsection{Objective Evaluation Protocol}
\label{sec:appendix_objective_evaluation}

\paragraph{Spectral Metrics.}
LSD, LF-LSD, and HF-LSD are computed from the final 48-kHz waveforms.
Let $X_{t,k}$ and $\widehat{X}_{t,k}$ denote the complex STFT
coefficients of the reference and generated waveforms at time frame $t$
and frequency bin $k$, respectively. We first compute their log-power
spectra as:
\begin{equation}
    Y_{t,k}
    =
    \log_{10}
    \left(
        \max\left(|X_{t,k}|^2,\epsilon\right)
    \right),
\end{equation}
where $\epsilon=10^{-8}$. The STFT uses a 2,048-sample Hann window with
a hop size of 512 samples.
For a frequency-bin set $\mathcal{K}$, the corresponding spectral distance
is defined as: 
\begin{equation}
    d(\mathcal{K})
    =
    \frac{1}{T}
    \sum_{t=1}^{T}
    \sqrt{
        \frac{1}{|\mathcal{K}|}
        \sum_{k\in\mathcal{K}}
        \left(
            \widehat{Y}_{t,k}-Y_{t,k}
        \right)^2
    }.
\end{equation}
Given the input cutoff frequency $f_c$, LSD is computed over the full
frequency range from 0 to 24~kHz. LF-LSD is computed over frequency bins
at or below $f_c$, whereas HF-LSD is computed over frequency bins above
$f_c$ and up to 24~kHz. Reference and generated signals are converted to mono before evaluation.
When their lengths differ, the longer signal is trimmed to the length of
the shorter one, without additional temporal alignment.

\paragraph{Perceptual Quality and Intelligibility Metrics.}
We report the overall quality score and coloration dimension predicted by
NISQA%
\footnote{\url{https://github.com/gabrielmittag/NISQA}}
as NISQA and COL, respectively. Speech intelligibility is evaluated using
the standard STOI metric. All metrics are computed utterance-wise and then
averaged over the complete evaluation set.

\subsection{Subjective Evaluation Protocol}
\label{sec:appendix_subjective_evaluation}

We conduct a listening test on QuestionPro to evaluate the overall quality
of the bandwidth-extended speech. The test covers five input sampling rates
(8, 12, 16, 20, and 24~kHz); for each rate, five utterances are selected and
the outputs of five systems are rated, with the original 48-kHz recordings
included as references. Fifteen university students (11 native English
speakers and 4 advanced L2 speakers) participate, each rating 130 samples
after being instructed to use headphones and set a comfortable volume. The
95\% confidence intervals are computed over the 75 ratings (15 listeners
$\times$ 5 utterances) per system and input rate.

\begin{figure}[!t]
    \centering
    \includegraphics[width=\linewidth]{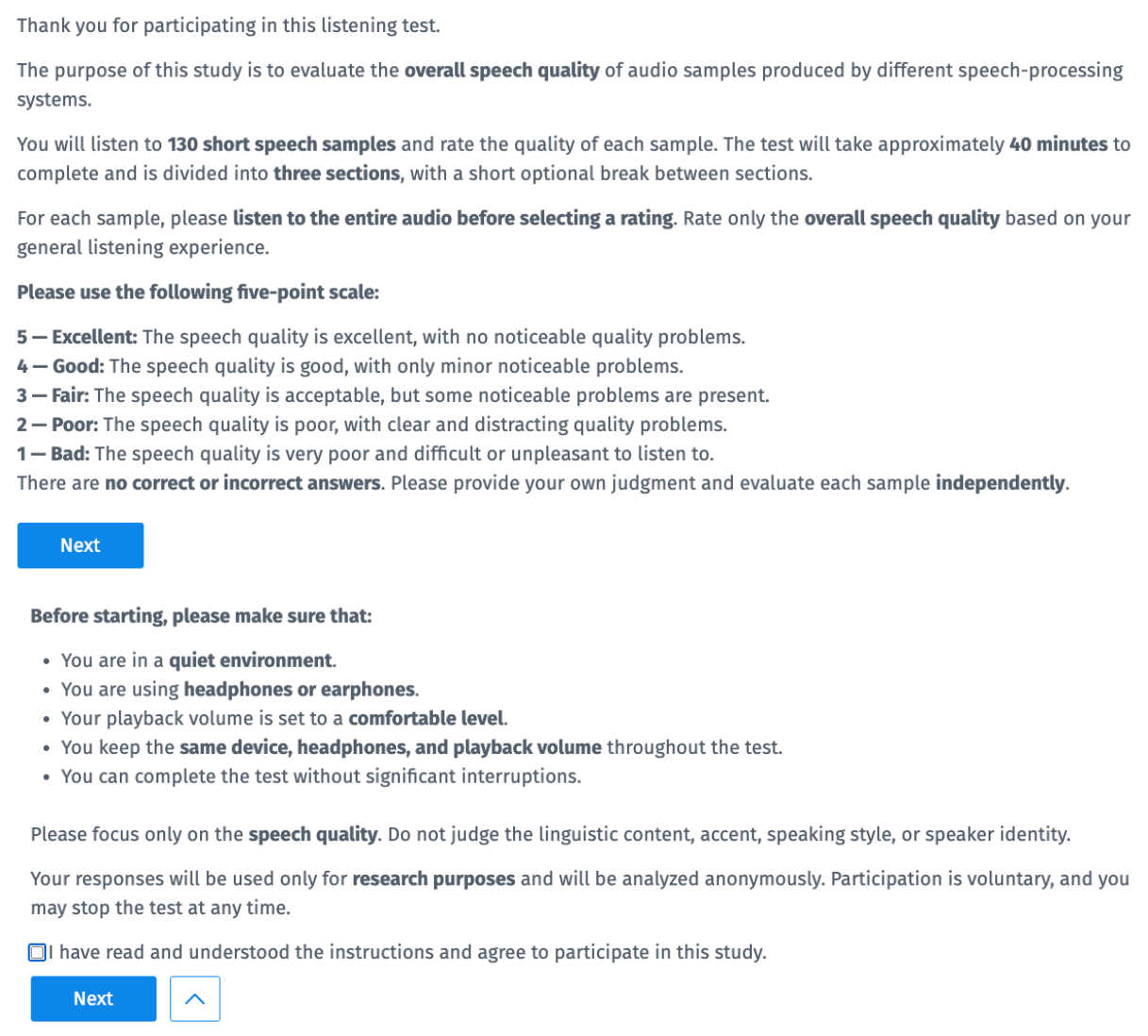}
    \caption{Instructions for the subjective listening test.}
    \label{fig:listening_test_instruction}
\end{figure}

\begin{figure}[!t]
    \centering
    \includegraphics[width=\linewidth]{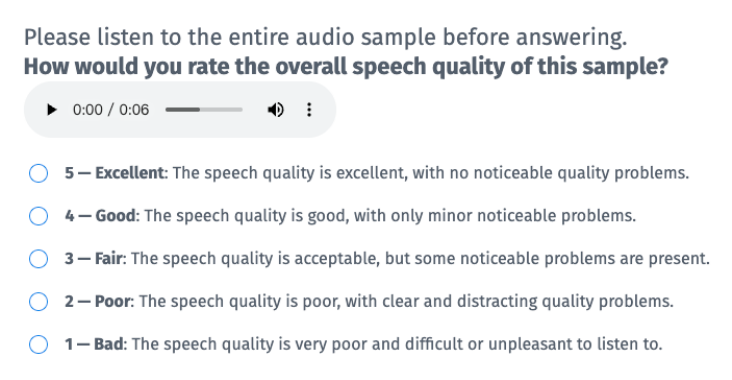}
    \caption{Interface for rating overall speech quality.}
    \label{fig:listening_test_rating}
\end{figure}

The test takes approximately 40 minutes and is divided into three sections,
with optional breaks between sections. Participants are instructed to conduct
the test in a quiet environment, use headphones or earphones, maintain a
comfortable and consistent playback volume, and listen to each complete
sample before assigning a score. Each sample is evaluated independently using
a five-point overall-speech-quality scale: 5 (\emph{Excellent}), 4
(\emph{Good}), 3 (\emph{Fair}), 2 (\emph{Poor}), and 1 (\emph{Bad}).
Participants are asked to focus only on speech quality and disregard the
linguistic content, accent, speaking style, and speaker identity. The final
score of each system is obtained by averaging the ratings across utterances
and listeners. The listening-test instructions and sample-rating interface are shown in
Figure~\ref{fig:listening_test_instruction} and
\ref{fig:listening_test_rating}, respectively.

\section{Additional Experimental Analyses}
\label{sec:appendix_experiments}
\subsection{Cutoff-Sampling Curriculum Analysis}
\label{sec:curriculum_analysis}

\begin{figure}[!t]
    \centering
    \includegraphics[width=\linewidth]{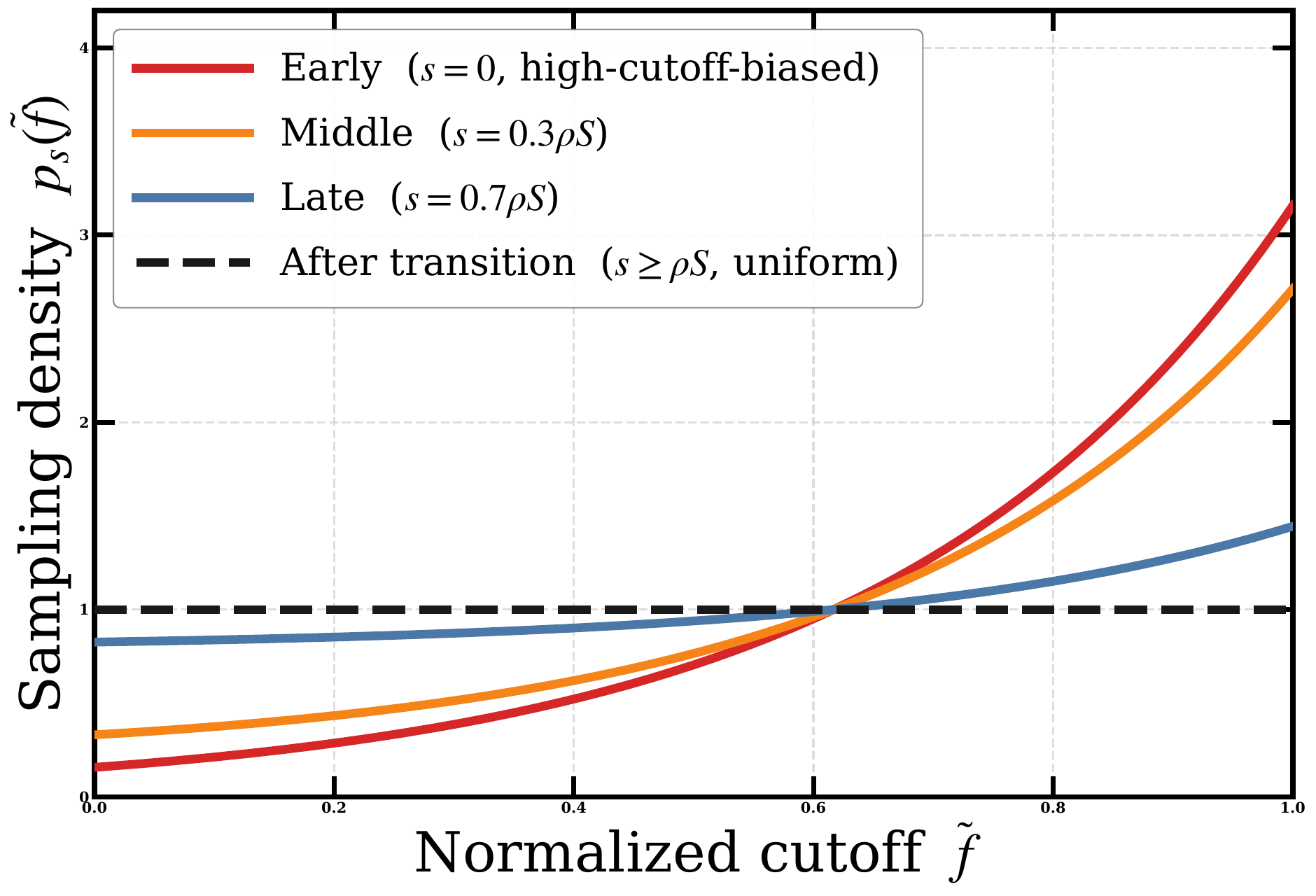}
    \caption{
    Illustration of the Easy-to-Balanced cutoff-sampling curriculum.
    Early training emphasizes larger normalized cutoffs, corresponding to
    easier inputs with narrower missing bands. The sampling density gradually
    transitions to a continuous uniform distribution, which is used after
    $s \geq \rho S$.
    }
    \label{fig:cutoff_curriculum}
\end{figure}

\paragraph{Cutoff-Sampling Curriculum.}
Figure~\ref{fig:cutoff_curriculum} illustrates the proposed
Easy-to-Balanced cutoff-sampling schedule. At the beginning of training,
the sampling density is biased toward larger normalized cutoffs, which
correspond to easier bandwidth-extension conditions with a smaller missing
frequency region. As training proceeds, the distribution gradually
interpolates toward the continuous uniform distribution over the full
cutoff range. After the transition stage ends at $s=\rho S$, the model is
trained using uniform continuous cutoff sampling. This schedule provides a
smooth progression from easier to more underdetermined bandwidth
conditions, while ultimately preserving balanced coverage of the entire
training range.

\paragraph{Irregular-Bandwidth Evaluation.}
\begin{table}[t]
\centering
\caption{
Ablation study of cutoff-sampling strategies on VCTK. Input sampling rates
of 3, 6, 12, and 15~kHz correspond to Nyquist cutoffs of 1.5, 3, 6, and
7.5~kHz, respectively. The output sampling rate is 48~kHz. The best and
second-best results within each input setting are highlighted in
\textbf{bold} and \underline{underlined}, respectively. Avg. denotes the
macro-average over the four input sampling rates.
}
\label{tab:ir_curriculum}

\scriptsize
\setlength{\tabcolsep}{2.4pt}
\renewcommand{\arraystretch}{1.02}

\resizebox{\columnwidth}{!}{
\begin{tabular}{clccccc}
\toprule
\textbf{Input SR}
& \textbf{Sampling strategy}
& \textbf{LSD} $\downarrow$
& \textbf{HF-LSD} $\downarrow$
& \textbf{NISQA} $\uparrow$
& \textbf{COL} $\uparrow$
& \textbf{STOI} $\uparrow$ \\
\midrule

\multirow{3}{*}{3 kHz}
& Uniform (discrete)
& 1.280
& 1.321
& 3.511
& \underline{3.602}
& 0.8454 \\

& Uniform (continuous)
& \underline{1.265}
& \underline{1.305}
& \underline{3.574}
& 3.464
& \underline{0.8485} \\

& Easy-to-Balanced
& \textbf{1.228}
& \textbf{1.261}
& \textbf{3.671}
& \textbf{3.702}
& \textbf{0.8523} \\
\midrule

\multirow{3}{*}{6 kHz}
& Uniform (discrete)
& 1.185
& 1.249
& 3.709
& 3.654
& 0.9412 \\

& Uniform (continuous)
& \underline{1.156}
& \underline{1.218}
& \underline{3.888}
& \underline{3.674}
& \underline{0.9505} \\

& Easy-to-Balanced
& \textbf{1.123}
& \textbf{1.177}
& \textbf{4.060}
& \textbf{4.002}
& \textbf{0.9575} \\
\midrule

\multirow{3}{*}{12 kHz}
& Uniform (discrete)
& 1.152
& 1.210
& 3.894
& 3.825
& 0.9904 \\

& Uniform (continuous)
& \underline{1.021}
& \underline{1.102}
& \underline{3.922}
& \textbf{3.954}
& \textbf{0.9985} \\

& Easy-to-Balanced
& \textbf{1.002}
& \textbf{1.093}
& \textbf{3.987}
& \underline{3.920}
& \underline{0.9968} \\
\midrule

\multirow{3}{*}{15 kHz}
& Uniform (discrete)
& 1.019
& 1.123
& 3.861
& \underline{3.956}
& 0.9905 \\

& Uniform (continuous)
& \underline{0.993}
& \underline{1.092}
& \textbf{4.034}
& \textbf{4.009}
& \underline{0.9975} \\

& Easy-to-Balanced
& \textbf{0.975}
& \textbf{1.085}
& \underline{3.975}
& 3.894
& \textbf{0.9987} \\
\midrule

\multirow{3}{*}{Avg.}
& Uniform (discrete)
& 1.159
& 1.226
& 3.744
& 3.759
& 0.9419 \\

& Uniform (continuous)
& \underline{1.109}
& \underline{1.179}
& \underline{3.855}
& \underline{3.775}
& \underline{0.9488} \\

& Easy-to-Balanced
& \textbf{1.082}
& \textbf{1.154}
& \textbf{3.923}
& \textbf{3.880}
& \textbf{0.9513} \\

\bottomrule
\end{tabular}
}
\end{table}
Table~\ref{tab:ir_curriculum} extends the cutoff-sampling comparison to
irregular input sampling rates. Unlike the regular-bandwidth results, where
Continuous Uniform does not consistently outperform Discrete Uniform,
Continuous Uniform achieves lower LSD and HF-LSD across all irregular
settings. This contrast highlights the importance of exposing the model to
continuously varying cutoffs when generalizing beyond the discrete training
points. Easy-to-Balanced further obtains the lowest LSD and HF-LSD at every
evaluated bandwidth, showing that its advantage cannot be attributed solely
to continuous cutoff coverage. The gains are particularly evident at the
more challenging 3- and 6-kHz inputs, where it also achieves the best
perceptual and intelligibility scores. Although Continuous Uniform is
slightly better on several perceptual metrics at higher input bandwidths,
Easy-to-Balanced achieves the best average result across all reported
metrics.

\begin{figure*}[!t]
    \centering
    \begin{subfigure}[t]{0.49\textwidth}
        \centering
        \includegraphics[width=\linewidth]
        {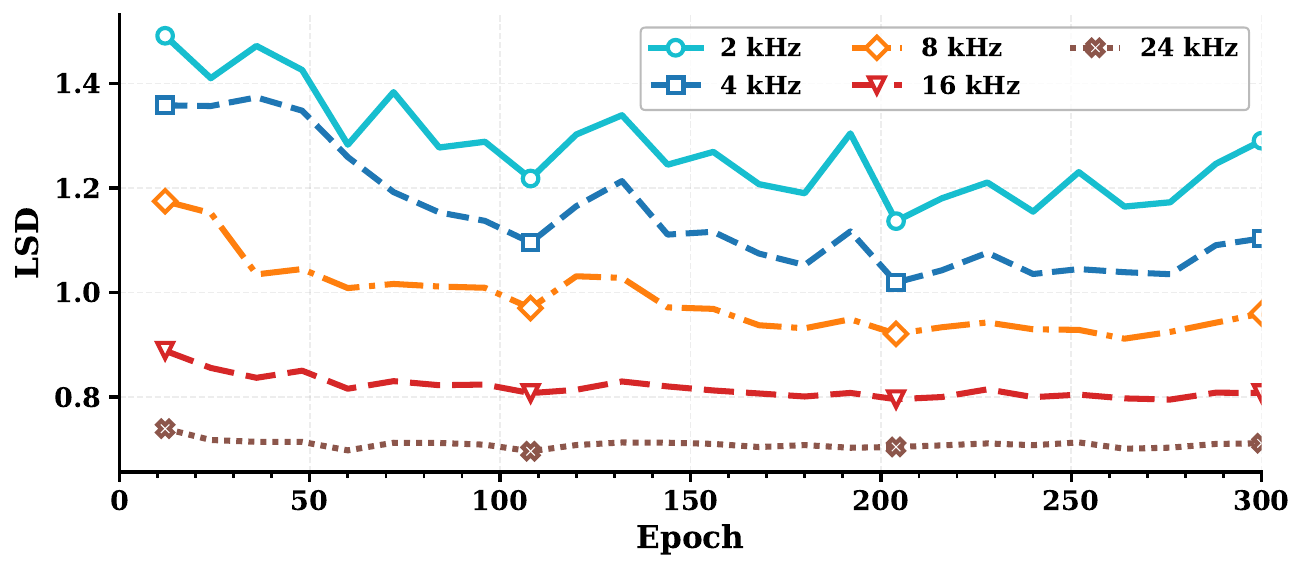}
        \caption{Absolute LSD.}
        \label{fig:easy_to_balanced_progress_abs}
    \end{subfigure}
    \hfill
    \begin{subfigure}[t]{0.49\textwidth}
        \centering
        \includegraphics[width=\linewidth]
        {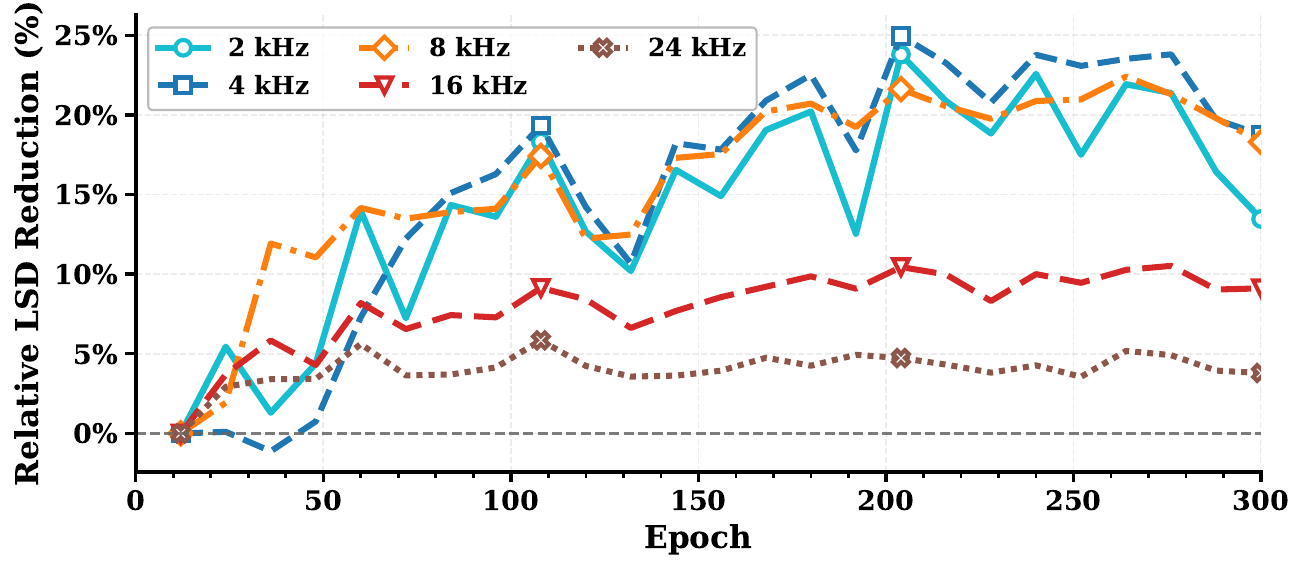}
        \caption{Relative LSD reduction.}
        \label{fig:easy_to_balanced_progress_rel}
    \end{subfigure}
    \caption{
    Stage-1 training progress under Easy-to-Balanced cutoff sampling across
    different input sampling rates. Relative reduction is measured with
    respect to the first evaluation checkpoint.
    }
    \label{fig:easy_to_balanced_progress}
\end{figure*}

\paragraph{Bandwidth-wise Effects of Easy-to-Balanced Training.}
To examine how different bandwidth conditions evolve under
Easy-to-Balanced training, we track the full-band LSD throughout Stage-1
training at input sampling rates of 2, 4, 8, 16, and 24~kHz.
Figure~\ref{fig:easy_to_balanced_progress} reports both the absolute LSD
and its relative reduction with respect to the first evaluation checkpoint.
For an input sampling rate $s$, the relative LSD reduction at epoch $e$ is
defined as
\begin{equation}
    R_{e}^{(s)}
    =
    \frac{
        \operatorname{LSD}_{e_0}^{(s)}
        -
        \operatorname{LSD}_{e}^{(s)}
    }{
        \operatorname{LSD}_{e_0}^{(s)}
    }
    \times 100\%,
    \label{eq:relative_lsd_reduction}
\end{equation}
where $e_0$ denotes the first evaluation checkpoint.

As shown in Figure~\ref{fig:easy_to_balanced_progress_abs}, lower input
sampling rates consistently result in higher absolute LSD, reflecting the
greater difficulty of reconstructing a wider missing frequency region.
After accounting for these different initial error levels,
Figure~\ref{fig:easy_to_balanced_progress_rel} shows that the challenging
2-, 4-, and 8-kHz conditions achieve larger overall relative reductions
than the 16- and 24-kHz conditions. The high-bandwidth conditions remain
comparatively stable, while the low-bandwidth curves exhibit larger
improvements together with some local fluctuations. This behavior is
consistent with the intended curriculum: training begins with relatively
informative inputs and progressively increases exposure to more severely
underdetermined bandwidth conditions before reaching a balanced continuous
cutoff distribution.

\paragraph{Hyperparameter sensitivity.}
To assess how sensitive the curriculum is to its two hyperparameters,
Table~\ref{tab:e2b_sweep} varies $\rho$ with $\beta{=}4$ and $\beta$ with
$\rho{=}0.7$. Every configuration improves NISQA and COL over continuous
uniform sampling, and all but one reduce LSD, whereas STOI shows no
consistent gain. Performance is stable for $\rho\in\{0.7,0.8\}$ and
degrades when annealing is too fast or too slow, or when the bias is too
weak or too strong, consistent with a balance between early training on
narrow missing bands and balanced coverage of all cutoffs. Discrete
uniform sampling remains competitive at the four regular rates it is
trained on but generalizes worse to irregular rates
as shown in Table~\ref{tab:ir_curriculum}. Because the sweep uses the test
speakers, we regard it as a sensitivity analysis rather than as
hyperparameter selection.
\begin{table}[t]
\centering
\caption{
Sensitivity of Easy-to-Balanced sampling to $\rho$ and $\beta$ on VCTK,
averaged over 2, 4, 8, and 16~kHz inputs. $^\dagger$Default setting. The
best and second-best results are highlighted in \textbf{bold} and
\underline{underlined}, respectively.
}
\label{tab:e2b_sweep}

\scriptsize
\setlength{\tabcolsep}{2.4pt}
\renewcommand{\arraystretch}{1.02}

\resizebox{\columnwidth}{!}{
\begin{tabular}{lcccccc}
\toprule
\textbf{Sampling strategy}
& $\boldsymbol{\rho}$
& $\boldsymbol{\beta}$
& \textbf{LSD} $\downarrow$
& \textbf{NISQA} $\uparrow$
& \textbf{COL} $\uparrow$
& \textbf{STOI} $\uparrow$ \\
\midrule

Uniform (discrete)
& -- & --
& 1.143
& 3.735
& 3.664
& \underline{0.9341} \\

Uniform (continuous)
& -- & --
& 1.183
& 3.545
& 3.527
& 0.9265 \\
\midrule

\multirow{8}{*}{Easy-to-Balanced}
& 0.5 & 4
& 1.188
& 3.630
& 3.693
& 0.9197 \\

& 0.6 & 4
& 1.170
& 3.639
& 3.659
& 0.9226 \\

& 0.7$^\dagger$ & 4
& \textbf{1.122}
& \underline{3.778}
& \textbf{3.801}
& \textbf{0.9358} \\

& 0.8 & 4
& \textbf{1.122}
& \textbf{3.789}
& \underline{3.788}
& 0.9336 \\

& 0.9 & 4
& 1.140
& 3.715
& 3.729
& 0.9251 \\

& 1.0 & 4
& 1.138
& 3.752
& 3.721
& 0.9253 \\
\cmidrule(lr){2-7}

& 0.7 & 2
& 1.155
& 3.697
& 3.616
& 0.9340 \\

& 0.7 & 8
& \underline{1.137}
& 3.701
& 3.722
& 0.9225 \\

\bottomrule
\end{tabular}
}
\end{table}

\subsection{Component Ablation across Input Bandwidths}
\label{sec:bandwidth_ablation}
\begin{table*}[!t]
\centering
\caption{
Component ablation across different input sampling rates on VCTK.
The corresponding Nyquist cutoffs are half of the input sampling rates.
w/o Stage 2 denotes the Stage-1 flow model without adversarial refinement.
The best and second-best results within each input setting are highlighted
in \textbf{bold} and \underline{underlined}, respectively.
}
\label{tab:ablation_across_bandwidths}

\scriptsize
\setlength{\tabcolsep}{2.0pt}
\renewcommand{\arraystretch}{1.05}

\resizebox{\textwidth}{!}{
\begin{tabular}{l*{4}{ccccc}}
\toprule

\multirow{2}{*}{\textbf{Variant}}
& \multicolumn{5}{c}{\textbf{Input SR: 2 kHz}}
& \multicolumn{5}{c}{\textbf{Input SR: 4 kHz}}
& \multicolumn{5}{c}{\textbf{Input SR: 8 kHz}}
& \multicolumn{5}{c}{\textbf{Input SR: 16 kHz}} \\

\cmidrule(lr){2-6}
\cmidrule(lr){7-11}
\cmidrule(lr){12-16}
\cmidrule(lr){17-21}

& LSD$\downarrow$
& HF-LSD$\downarrow$
& NISQA$\uparrow$
& COL$\uparrow$
& STOI$\uparrow$

& LSD$\downarrow$
& HF-LSD$\downarrow$
& NISQA$\uparrow$
& COL$\uparrow$
& STOI$\uparrow$

& LSD$\downarrow$
& HF-LSD$\downarrow$
& NISQA$\uparrow$
& COL$\uparrow$
& STOI$\uparrow$

& LSD$\downarrow$
& HF-LSD$\downarrow$
& NISQA$\uparrow$
& COL$\uparrow$
& STOI$\uparrow$ \\
\midrule

AnyBand (Ours)
& \textbf{1.248}
& \textbf{1.269}
& \textbf{3.125}
& \textbf{3.419}
& \textbf{0.8214}
& \underline{1.180}
& \underline{1.219}
& \textbf{4.038}
& \textbf{3.966}
& \textbf{0.9356}
& \underline{1.086}
& 1.155
& \textbf{4.014}
& \textbf{3.983}
& \underline{0.9870}
& \textbf{0.974}
& \underline{1.092}
& \textbf{3.936}
& \textbf{3.837}
& \underline{0.9992} \\
\midrule

\quad $-$ w/o Freq. Modules
& 1.327
& 1.352
& 2.882
& 2.843
& 0.7940
& 1.215
& 1.260
& 3.920
& 3.827
& 0.9140
& 1.090
& \underline{1.141}
& 3.918
& 3.856
& 0.9850
& 1.038
& 1.159
& 3.772
& 3.656
& 0.9978 \\

\quad $-$ w/o F0/UV
& 1.332
& 1.359
& 2.905
& 2.980
& 0.8067
& 1.253
& 1.299
& 3.803
& 3.707
& 0.9091
& 1.137
& 1.204
& 3.848
& 3.820
& 0.9830
& 1.035
& 1.125
& 3.758
& 3.698
& 0.9985 \\

\quad $-$ w/o $\mathcal{D}_{\mathrm{spec}}$
& 1.354
& 1.372
& 2.543
& 2.790
& 0.7852
& 1.229
& 1.278
& 3.825
& 3.843
& 0.8966
& 1.156
& 1.218
& 3.782
& 3.609
& 0.9725
& 1.124
& 1.206
& 3.673
& 3.628
& 0.9935 \\

\quad $-$ w/o $\mathcal{D}_{\mathrm{cross}}$
& 1.295
& 1.317
& \underline{3.012}
& 3.286
& \underline{0.8133}
& \textbf{1.157}
& \textbf{1.181}
& \underline{4.027}
& \underline{3.940}
& 0.9204
& 1.122
& 1.165
& \underline{3.955}
& \underline{3.896}
& \textbf{0.9872}
& \underline{0.998}
& \textbf{1.076}
& \underline{3.847}
& \underline{3.752}
& 0.9968 \\

\quad $-$ w/o $\mathcal{D}_{\mathrm{harm}}$
& \underline{1.280}
& \underline{1.302}
& 3.000
& \underline{3.322}
& 0.8095
& 1.193
& 1.235
& 3.882
& 3.756
& \underline{0.9310}
& \textbf{1.069}
& \textbf{1.127}
& 3.894
& 3.796
& 0.9865
& 1.012
& 1.098
& 3.760
& 3.735
& \textbf{0.9996} \\

\quad $-$ w/o Stage 2
& 1.357
& 1.379
& 2.511
& 2.752
& 0.7855
& 1.233
& 1.275
& 3.818
& 3.767
& 0.8954
& 1.163
& 1.234
& 3.776
& 3.633
& 0.9730
& 1.123
& 1.217
& 3.678
& 3.602
& 0.9932 \\

\bottomrule
\end{tabular}
}
\end{table*}

Table~\ref{tab:ablation_across_bandwidths} extends the component ablation
to multiple input sampling rates (2, 4, 8, and 16 kHz). Removing either the frequency-aware
modules or the F0/VUV conditioning generally degrades spectral reconstruction,
perceptual quality, and intelligibility, with the effects being most
pronounced under severely bandwidth-limited inputs. This indicates that
explicit frequency-wise modeling and pitch--voicing cues become particularly
important when only a small portion of the original spectrum is observed.
Their influence becomes less pronounced as the available bandwidth
increases, although the complete model provides a more balanced performance
across the reported metrics.
To assess the adversarial refinement stage as a whole,
Table~\ref{tab:ablation_across_bandwidths} also includes the Stage-1 flow model before
refinement. Stage~2 improves all five metrics at every input rate. Since
this comparison also includes the additional training of Stage~2, it
supports the refinement stage as a whole rather than isolating the
endpoint-focused design.

Among the adversarial components, removing
$\mathcal{D}_{\mathrm{spec}}$ results in the most consistent degradation
across input bandwidths, confirming the importance of multi-scale spectral
realism during adversarial refinement. The contributions of
$\mathcal{D}_{\mathrm{cross}}$ and $\mathcal{D}_{\mathrm{harm}}$ are more
dependent on the bandwidth and evaluation metric. Removing either component
occasionally improves an individual metric at higher input sampling rates,
but generally weakens perceptual quality and performance under more
challenging low-bandwidth conditions. Overall, the complete AnyBand model
achieves the best average performance across all reported metrics, suggesting
that the proposed components provide complementary benefits across different
degrees of bandwidth limitation. 

In addition, removing the frequency modules changes both
how the model relates frequency bins within a frame and its size, so the
gain of the full model could in principle come from added capacity. To
separate the two, we build a parameter-matched control. The frequency
encoder and decoder treat each frame as a sequence of mel-bin tokens and
apply two frequency-axis Transformer blocks each, letting observed and
missing bins attend to one another, at a cost of only 0.09M parameters over
the frame-wise linear projections they replace. The control removes both
modules and instead inserts four frame-wise residual MLP blocks of matched
size, which cannot attend across frequency bins. As
Table~\ref{tab:capacity} shows, the control changes performance only
marginally relative to the variant without frequency modules, whereas the
full model improves all four metrics; the gain therefore comes from
frequency-axis attention rather than added parameters. The control matches
parameters but not computation.
\begin{table}[t]
\centering
\caption{
Parameter-matched control for the frequency modules on VCTK, averaged over
2, 4, 8, and 16~kHz inputs. Gen.\ denotes generator parameters, excluding
the vocoder. The best and second-best results are highlighted in
\textbf{bold} and \underline{underlined}, respectively.
}
\label{tab:capacity}

\scriptsize
\setlength{\tabcolsep}{2.4pt}
\renewcommand{\arraystretch}{1.02}

\resizebox{\columnwidth}{!}{
\begin{tabular}{lccccc}
\toprule
\textbf{Model}
& \textbf{Gen. (M)}
& \textbf{LSD} $\downarrow$
& \textbf{NISQA} $\uparrow$
& \textbf{COL} $\uparrow$
& \textbf{STOI} $\uparrow$ \\
\midrule

w/o Freq. Modules
& 22.1681
& \underline{1.168}
& 3.623
& 3.546
& 0.9227 \\

MLP control
& 22.2542
& 1.170
& \underline{3.640}
& \underline{3.555}
& \underline{0.9253} \\

AnyBand
& 22.2540
& \textbf{1.122}
& \textbf{3.778}
& \textbf{3.801}
& \textbf{0.9358} \\

\bottomrule
\end{tabular}
}
\end{table}

\subsection{Prediction Parameterization}
\label{sec:prediction_parameterization}
\begin{table}[t]
\centering
\caption{
Comparison of $x$- and $v$-prediction under different numbers of sampling
steps on VCTK with a 16-kHz input sampling rate (8-kHz Nyquist cutoff).
The best result under each sampling-step setting is highlighted in
\textbf{bold}.
}
\label{tab:xv_prediction}
\scriptsize
\setlength{\tabcolsep}{2.8pt}
\renewcommand{\arraystretch}{1.02}
\begin{tabular}{lcccccc}
\toprule
Prediction
& Steps
& LSD $\downarrow$
& HF-LSD $\downarrow$
& NISQA $\uparrow$
& COL $\uparrow$
& STOI $\uparrow$ \\
\midrule

$v$-prediction
& 32
& \textbf{1.065}
& \textbf{1.107}
& \textbf{4.060}
& \textbf{3.825}
& \textbf{0.9912} \\

$x$-prediction
& 32
& 1.075
& 1.194
& 3.883
& 3.796
& 0.9905 \\

\midrule

$v$-prediction
& 16
& \textbf{1.132}
& \textbf{1.224}
& 3.757
& \textbf{3.710}
& \textbf{0.9844} \\

$x$-prediction
& 16
& 1.235
& 1.301
& \textbf{3.764}
& 3.686
& 0.9726 \\

\bottomrule
\end{tabular}
\end{table}





\paragraph{Compared Parameterizations.}
We further compare \emph{$v$-prediction}, which directly estimates the flow
velocity~\cite{lipman2022flow}, with \emph{$x$-prediction}, which predicts
the clean spectrogram and converts it to velocity for flow matching and ODE
sampling, following JiT~\cite{li2026back}. The two variants use the same
probability path, velocity-space training objective, model architecture, and
sampling procedure, differing only in the quantity directly predicted by the
generator.

Under $v$-prediction, the generator directly estimates the velocity field:
\begin{equation}
    \widehat{\mathbf v}_{\theta}
    =
    \mathcal G_{\theta}
    \left(
        \mathbf M_t,
        \widetilde{\mathbf M},
        t,
        \mathbf c
    \right).
\end{equation}
Under $x$-prediction, the generator instead predicts the clean full-band
spectrogram:
\begin{equation}
    \widehat{\mathbf M}_{\theta}
    =
    \mathcal G_{\theta}
    \left(
        \mathbf M_t,
        \widetilde{\mathbf M},
        t,
        \mathbf c
    \right),
\end{equation}
which is converted into the corresponding velocity as: 
\begin{equation}
    \widehat{\mathbf v}_{\theta}
    =
    \frac{
        \widehat{\mathbf M}_{\theta}-\mathbf M_t
    }{\delta_t},
    \qquad
    \delta_t=\max(1-t,\delta).
\end{equation}
The resulting velocity is used for both the masked flow-matching objective
and ODE sampling. Although the two parameterizations are algebraically
equivalent under perfect prediction, they impose different prediction
targets on the generator. $v$-prediction directly estimates the vector field
required by the ODE solver, whereas $x$-prediction first estimates the clean
endpoint and then applies a time-dependent conversion to obtain the velocity.

\paragraph{Results.}
Table~\ref{tab:xv_prediction} compares the two parameterizations under
otherwise identical configurations. At 32 sampling steps, $v$-prediction
outperforms $x$-prediction across all reported metrics. At 16 steps,
$v$-prediction continues to achieve better LSD, HF-LSD, COL, and STOI,
whereas $x$-prediction obtains only a marginally higher NISQA score. Overall,
$v$-prediction provides more favorable performance under both standard and
reduced sampling budgets, and is therefore adopted in AnyBand.

\subsection{MOSNet Evaluation}
\begin{figure}[t]
\centering
\includegraphics[width=\columnwidth]{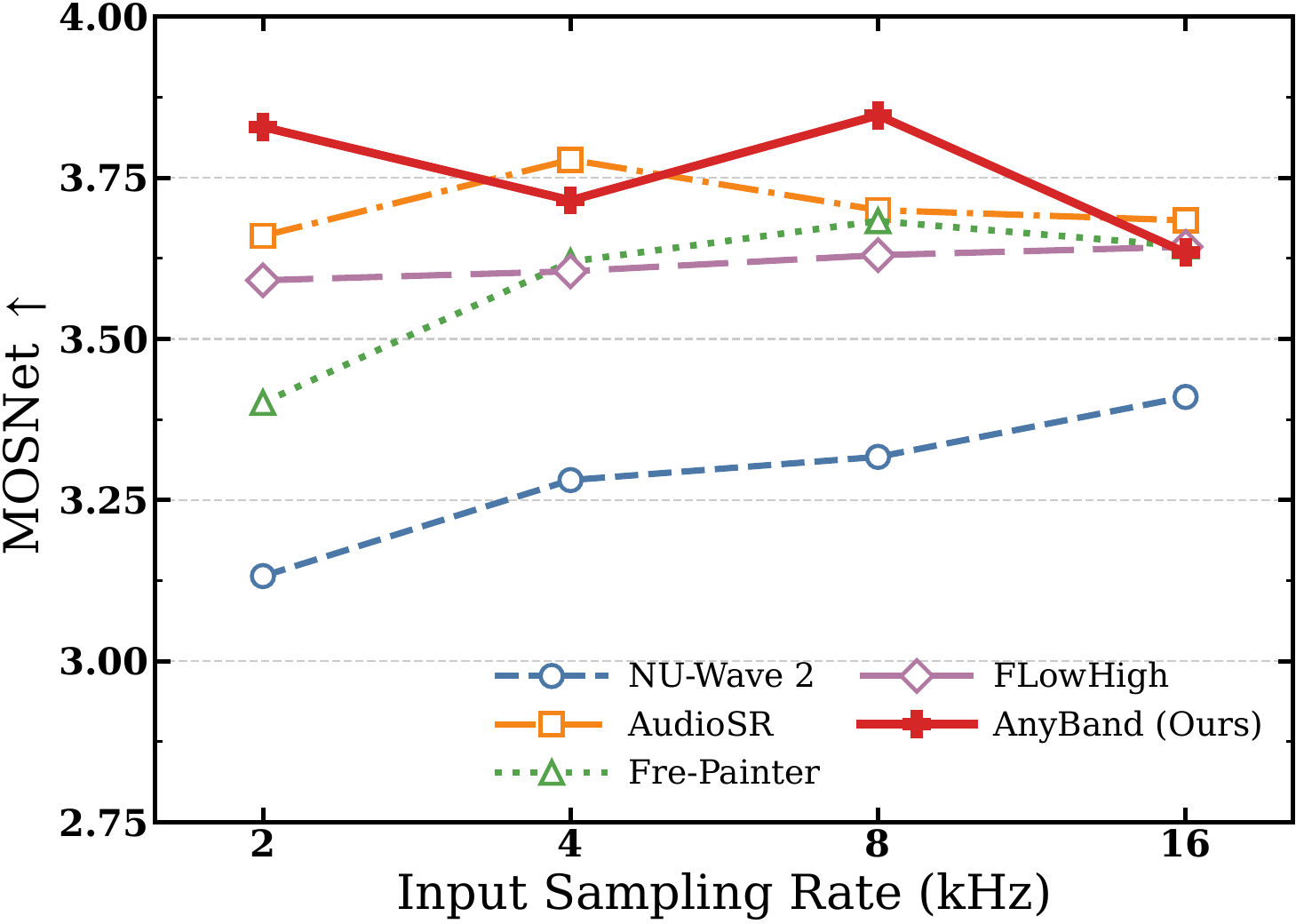}
\caption{MOSNet on VCTK across input sampling rates. MOSNet operates on
16~kHz audio and thus reflects only content below 8~kHz.}
\label{fig:mosnet}
\end{figure}

The listening test in Figure~\ref{fig:subjective_quality} covers 8--24~kHz inputs.
To also assess perceived quality for the most band-limited inputs,
Figure~\ref{fig:mosnet} reports MOSNet on VCTK at 2--16~kHz inputs.
AnyBand achieves the highest MOSNet on average and at 2 and 8~kHz, and
ranks second at 4~kHz. Because MOSNet operates on 16~kHz audio, it cannot
assess content above 8~kHz and serves only as a proxy for perceptual
quality; at 16~kHz inputs it sees only the observed band, so differences
there mainly reflect waveform synthesis.

\subsection{Inference Cost}
Table~\ref{tab:cost} compares model size and inference speed. AnyBand is
slower than NU-Wave~2, FLowHigh, and Fre-Painter but faster than AudioSR;
its cost is dominated by the 50-step sampler with classifier-free guidance,
whose quality--steps trade-off is shown in Figure~5.
\begin{table}[t]
\centering
\caption{
Model size and inference speed. Parameter counts of AnyBand and FLowHigh
include their vocoders. RTF is measured end to end on a single GPU (FP32,
batch size 1). The best and second-best RTF are highlighted in
\textbf{bold} and \underline{underlined}, respectively.
}
\label{tab:cost}

\scriptsize
\setlength{\tabcolsep}{4pt}
\renewcommand{\arraystretch}{1.02}

\resizebox{0.6\columnwidth}{!}{
\begin{tabular}{lcc}
\toprule
\textbf{Method}
& \textbf{Params (M)}
& \textbf{RTF} $\downarrow$ \\
\midrule

NU-Wave 2
& 1.71
& 0.2028 \\

AudioSR
& 1085.83
& 0.9580 \\

FLowHigh
& 49.40
& \underline{0.0161} \\

Fre-Painter
& 22.91
& \textbf{0.0083} \\
\midrule

AnyBand (Ours)
& 58.95
& 0.5139 \\

\bottomrule
\end{tabular}
}
\end{table}



\section{Qualitative Results}
\label{sec:appendix_qual}
\label{sec:qualitative_comparisons}

\begin{figure*}[!t]
    \centering
    \begin{subfigure}[t]{\linewidth}
        \centering
        \includegraphics[width=\linewidth]
        {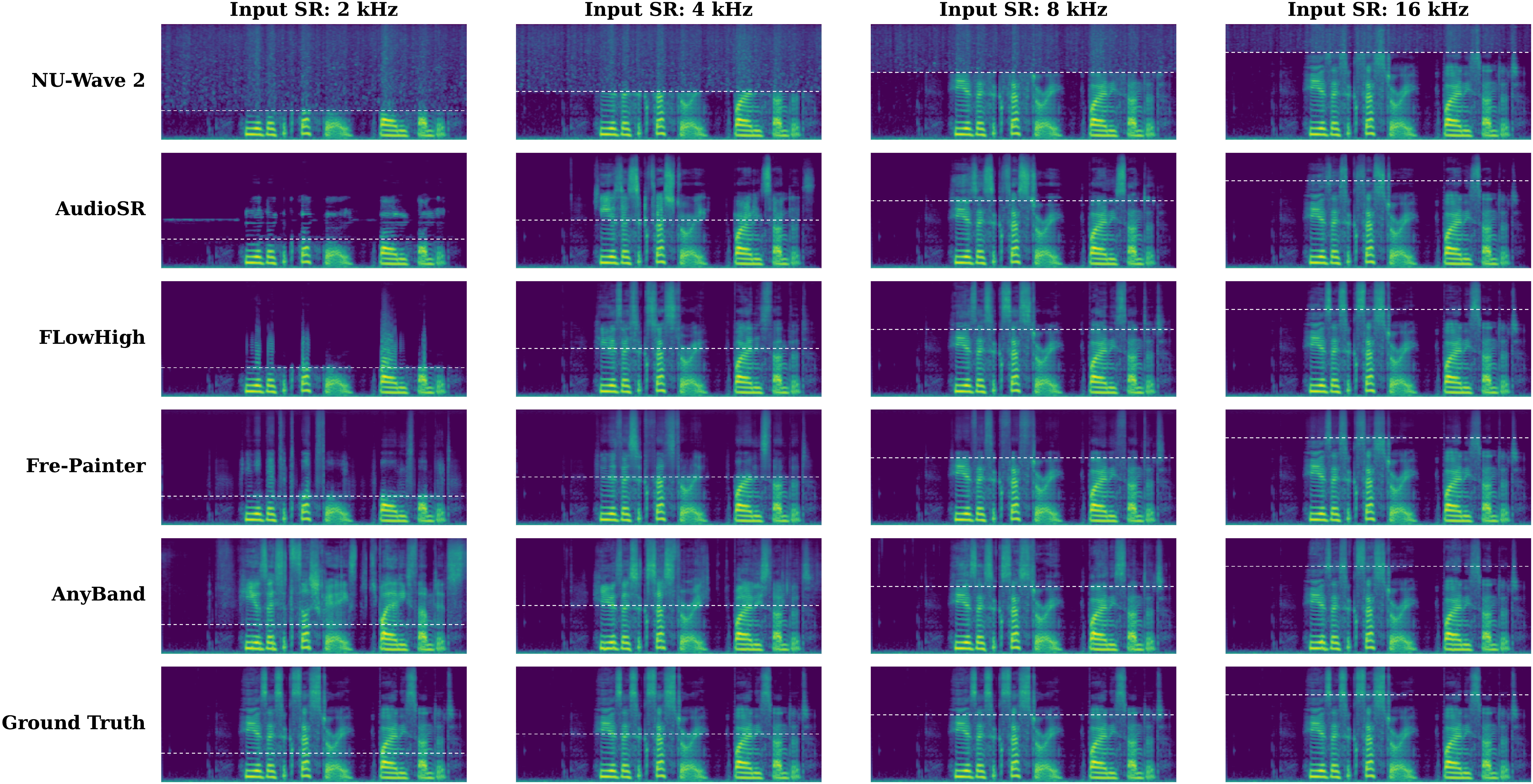}
        \caption{Standard input sampling rates.}
        \label{fig:qualitative_p225_standard}
    \end{subfigure}

    \vspace{1mm}

    \begin{subfigure}[t]{\linewidth}
        \centering
        \includegraphics[width=\linewidth]{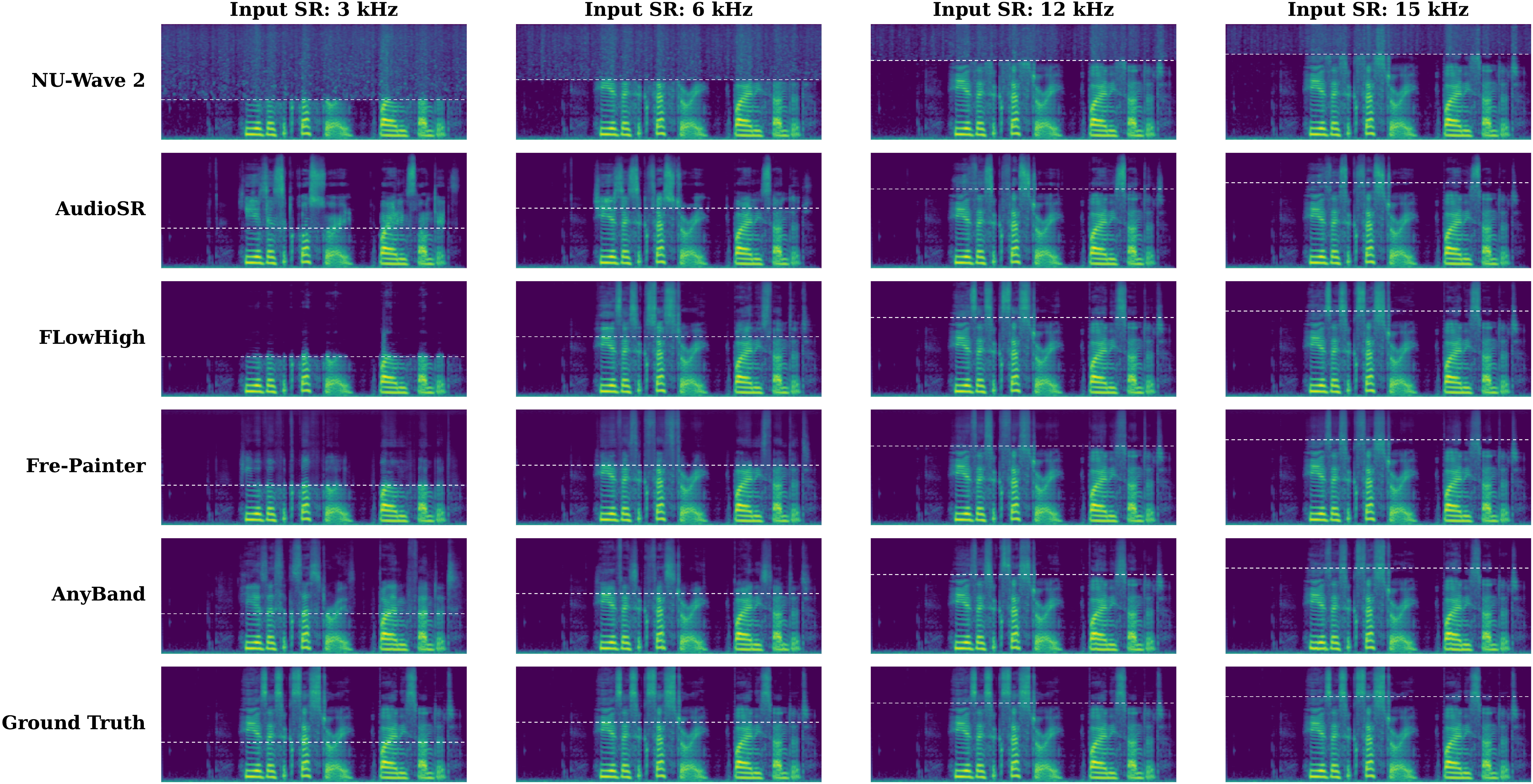}
        \caption{Irregular input sampling rates.}
        \label{fig:qualitative_p225_irregular}
    \end{subfigure}

    \caption{
    Qualitative mel-spectrogram comparisons for a representative utterance
    under standard and irregular input bandwidths. Horizontal dashed lines
    indicate the corresponding cutoff frequencies.
    }
    \label{fig:qualitative_p225}
\end{figure*}

\begin{figure*}[!t]
    \centering
    \begin{subfigure}[t]{\linewidth}
        \centering
        \includegraphics[width=\linewidth]{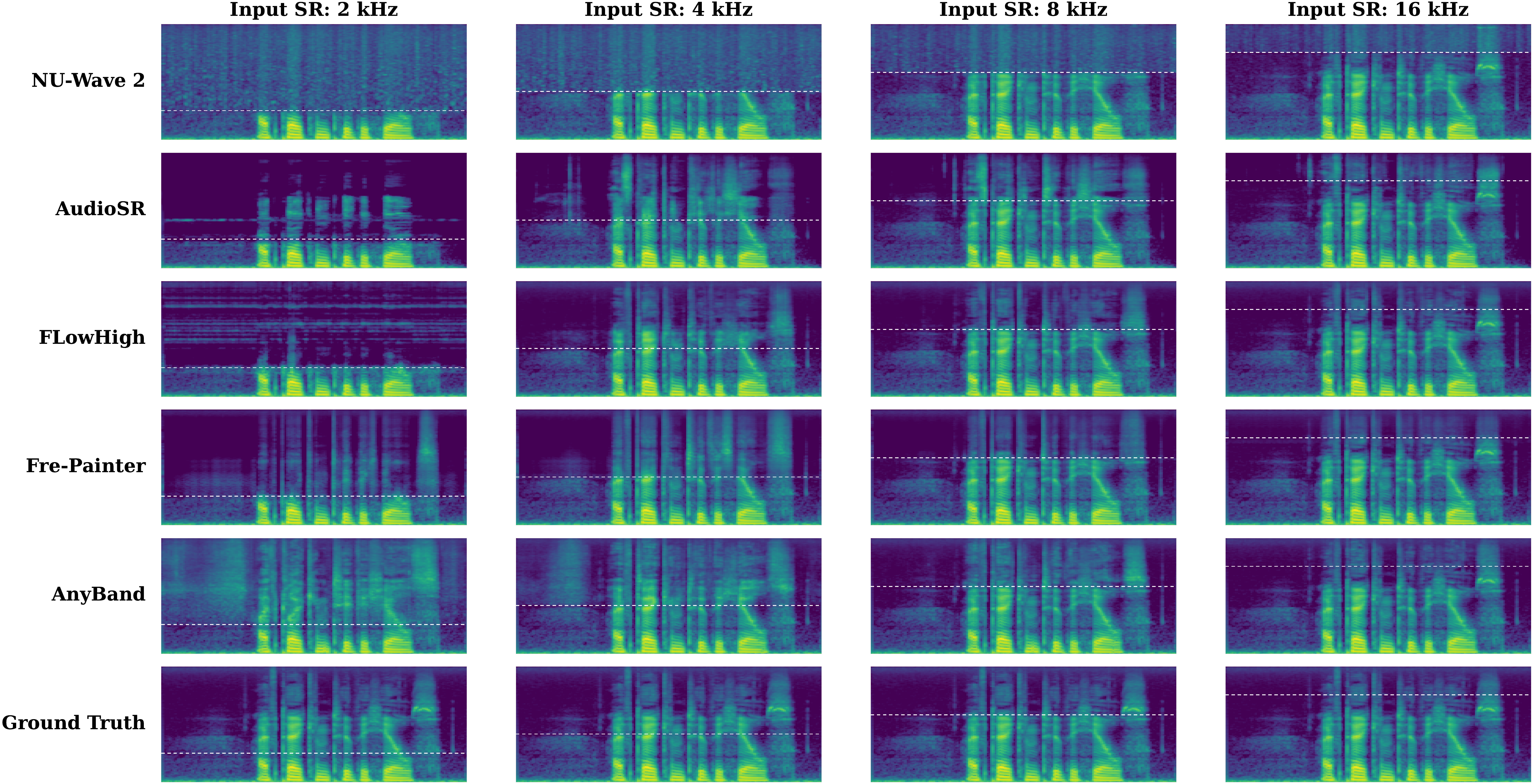}
        \caption{Standard input sampling rates.}
        \label{fig:qualitative_example_two_standard}
    \end{subfigure}

    \vspace{1mm}

    \begin{subfigure}[t]{\linewidth}
        \centering
        \includegraphics[width=\linewidth]{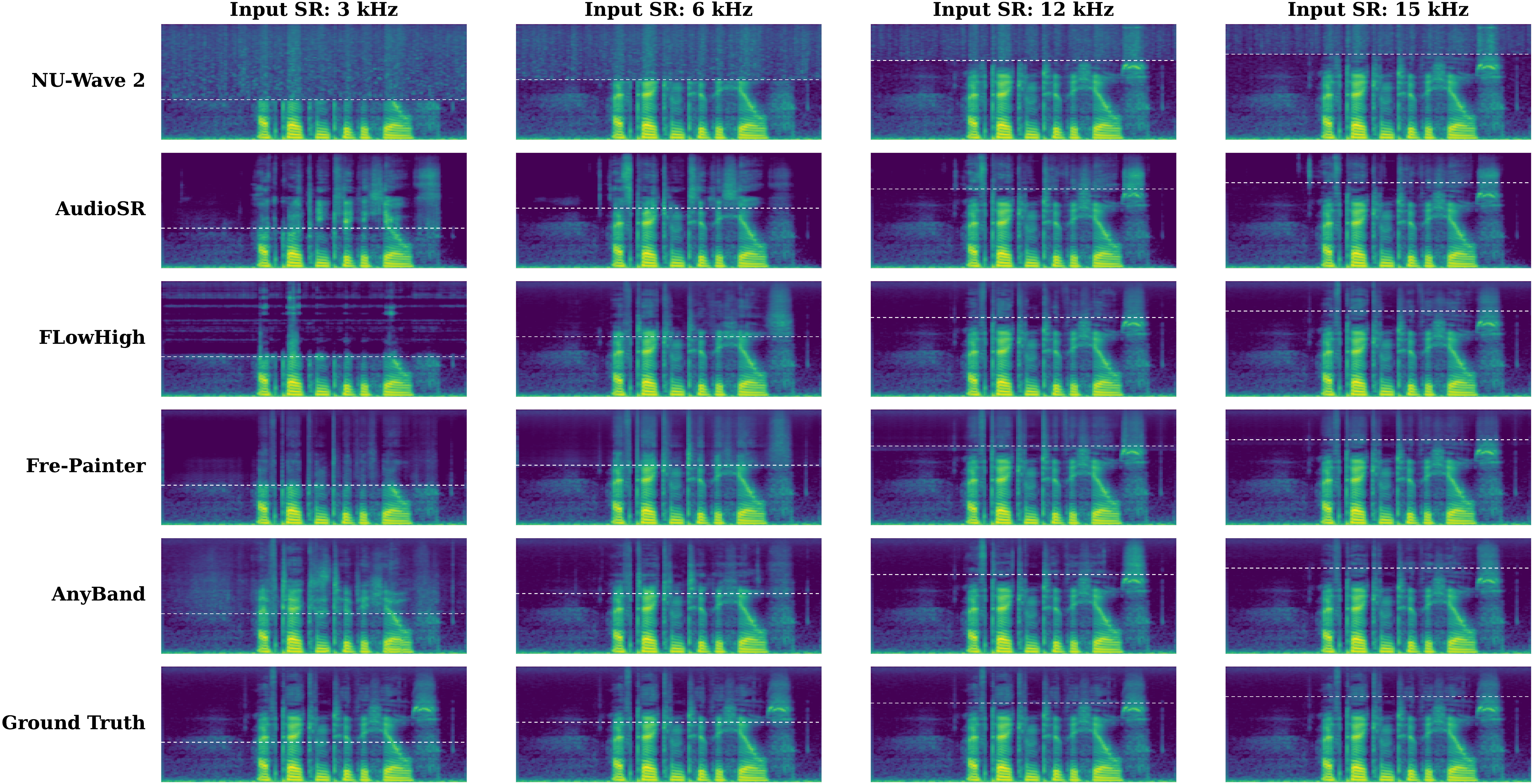}
        \caption{Irregular input sampling rates.}
        \label{fig:qualitative_example_two_irregular}
    \end{subfigure}

    \caption{
    Additional qualitative mel-spectrogram comparisons under standard and
    irregular input bandwidths.
    }
    \label{fig:qualitative_example_two}
\end{figure*}

Figures~\ref{fig:qualitative_p225} and
\ref{fig:qualitative_example_two} present qualitative mel-spectrogram
comparisons for two representative utterances under both standard and
irregular input bandwidths. The standard settings include input sampling
rates of 2, 4, 8, and 16~kHz, while the irregular settings include
3, 6, 12, and 15~kHz. Each column corresponds to one input sampling rate,
and the horizontal dashed line indicates the associated cutoff frequency.

The differences are most apparent under the severely bandwidth-limited
2--6~kHz inputs. NU-Wave 2 tends to produce diffuse high-frequency textures,
while AudioSR and FLowHigh occasionally generate sparse, discontinuous, or
banded structures above the cutoff. Fre-Painter recovers broader
high-frequency energy, but some fine-grained spectral structures remain
over-smoothed. In comparison, AnyBand generates more coherent
high-frequency patterns that remain temporally aligned with the observed
low-frequency content and transition more smoothly across the cutoff
boundary. Its reconstructed harmonic and broadband structures are also
visually closer to those of the ground-truth spectrograms.

As the available input bandwidth increases, the outputs of all methods
become more similar to the ground truth, although AnyBand continues to
preserve clearer high-frequency details and spectral continuity. Similar
behavior is observed at the irregular input sampling rates, indicating that
the learned spectral-continuation behavior is not restricted to the
standard bandwidth settings.


\end{document}